\documentclass[modern]{aastex63}

\usepackage{natbib,graphics,graphicx,mhchem,url,hyperref} 
\bibpunct{(}{)}{;}{a}{,}{,}

\newcommand{\GG}[1]{}
\usepackage[symbol]{footmisc}

\begin{document}

\title{Subsurface Thermophysical Properties of Europa's Leading and Trailing
  Hemispheres as Revealed by ALMA}

\correspondingauthor{Alexander E. Thelen}
\email{athelen@caltech.edu} 

\author{Alexander E. Thelen}
\affiliation{Division of Geological and Planetary Sciences, California
  Institute of Technology, Pasadena, CA 91125, USA}

\author{Katherine de Kleer}
\affiliation{Division of Geological and Planetary Sciences, California
  Institute of Technology, Pasadena, CA 91125, USA}

\author{Maria Camarca}
\affiliation{Division of Geological and Planetary Sciences, California
  Institute of Technology, Pasadena, CA 91125, USA}

\author{Alexander Akins}
\affiliation{Jet Propulsion Laboratory, California
  Institute of Technology, Pasadena, CA 91011, USA}

\author{Mark Gurwell}
\affiliation{Center for Astrophysics $|$ Harvard $\&$ Smithsonian, Cambridge, MA
  02138, USA}

\author{Bryan Butler}
\affiliation{National Radio Astronomy Observatory, Socorro, NM 87801, USA}

\author{Imke de Pater}
\affiliation{Department of Astronomy $\&$ Department of Earth and
Planetary Science, Campbell Hall 501, University of California,
Berkeley CA 94720, USA}

\begin{abstract}

We present best-fit values of porosity -- and the corresponding effective
thermal inertiae -- determined from three different depths in Europa's
near-subsurface ($\sim1-20$ cm). The porosity of the upper
$\sim20$ cm of Europa's subsurface varies between 75--50$\%$ ($\Gamma_{eff}\approx50-140$
J m$^{-2}$ K$^{-1}$ s$^{-1/2}$) on the leading
hemisphere and 50--40$\%$ ($\Gamma_{eff}\approx140-180$
J m$^{-2}$ K$^{-1}$ s$^{-1/2}$) on the trailing hemisphere.
 Residual maps
produced by comparison with these models reveal thermally anomalous
features that
cannot be reproduced by globally homogeneous porosity models. These regions
are compared to Europa's surface terrain and known compositional
variations. We find that some instances of warm thermal anomalies are co-located 
with known geographical or compositional features on both the leading
and trailing hemisphere; cool
temperature anomalies are well correlated with
  surfaces previously observed to contain pure, crystalline water ice and
  the expansive rays of Pwyll crater. Anomalous regions correspond to
locations with subsurface properties different from those of our best-fit
models, such as potentially elevated thermal inertia, decreased emissivity, or more
porous regolith. We also
find that ALMA observations at $\sim3$ mm sound below the
thermal skin depth of Europa ($\sim10-15$ cm) for a range of
porosity values, and thus do not exhibit features
indicative of diurnal variability or residuals similar to other
frequency bands. Future observations of Europa at higher angular resolution
may reveal additional locations of variable subsurface
thermophysical properties, while those at other wavelengths
will inform our understanding of the regolith compaction length and
the effects of external processes on the shallow subsurface.

\end{abstract}

\section{Introduction} \label{sec:intro}
\renewcommand{\thefootnote}{\arabic{footnote}}
The surface of Europa, the smallest of the Galilean Satellites, is
notable for its varied terrain units and hemispheric asymmetries --
particularly when compared to its sister moons that are resurfaced by
active volcanoes (Io), heavily cratered and relatively dark
(Callisto), or somewhat intermediary (Ganymede) (see the reviews in
\citealt{mcewen_04}, \citealt{greeley_04}, \citealt{moore_04},
  \citealt{pappalardo_04}, \citealt{de_pater_21a} and references therein). The presence of ridged plains, chaotic terrain (comprised of small, incoherent
ice latticework), hydrated salts and sulfuric compounds across Europa's icy surface indicates the
crust above its subsurface ocean may be relatively young and
tectonically active, while also being exogenically weathered
\citep{smith_79, mccord_98, zahnle_98, cooper_01, paranicas_01, zahnle_03,
  schenk_04, bierhaus_09, 
doggett_09, carlson_09}. Tidally locked and orbiting slower than
Jupiter's magnetic field, which is tied to the planet's rapid rotation
($\sim10$ hr), the trailing hemisphere (centered at 270$^{\circ}$
W) is constantly bombarded by charged particles and heavy ions
(e.g. S$^+$, O$^+$) sourced from Io's plasma torus that are entrained in
Jupiter's magnetic field \citep{paranicas_09}. The leading hemisphere (centered at 90$^{\circ}$
W) is exposed to the highest energy particles from the Jovian
magnetosphere \citep{nordheim_22} and micrometeorite gardening \citep{zahnle_98}, and presents a brighter, less altered surface. However,
the influence of exogenic erosion of the upper layers of
Europa's regolith and the slow resurfacing from beneath are
not easily related to latitude or longitude, resulting in a complex surface
whose composition and structure are influenced from both the
  subsurface ocean and exogenic environment 
  \citep{anderson_98, carr_98, pappalardo_99, kivelson_00}. Hydrated
  minerals and salts have been detected across the varied surface
terrain, possibly originating in the subsurface while radiolysis provides the formation of
sulfur-bearing species, hydrogen peroxide, carbon dioxide, among others \citep{mccord_98, carlson_05, carlson_09, brown_13, trumbo_19a, trumbo_19b,
  trumbo_22, trumbo_23, villanueva_23}.

As far back as the early 20$^{th}$ century questions regarding the hemispheric
dichotomy of Europa's surface brightness and properties began to
arise, initially from ground-based observations \citep{stebbins_27,
  stebbins_28}. These questions persisted into the 1970's based on further
ground-based observations and data from the
Pioneer flybys of Jupiter \citep{fimmel_74}.
Subsequently, the coloration, non-icy material composition, mixtures of
amorphous or crystalline ice, and weathering by Jovian
magnetospheric ions have been investigated in-depth using near-infrared through ultraviolet
wavelength 
instruments onboard the spacecraft venturing near
and into
the Jovian system. Initial observations from the Voyager spacecraft
determined differences in color, albedo, and water ice distributions between the
leading and trailing hemispheres \citep{pilcher_72, lucchitta_82,
  mcewen_86, spencer_87}, while subsequent Galileo measurements
revealed compositional and thermal variations using the 
Ultraviolet Spectrometer (UVS; \citealp{hendrix_98}), Near Infrared Mass Spectrometer
(NIMS; \citealp{carlson_96, mccord_98, hansen_04}), Photopolarimeter-radiometer
(PPR; \citealp{spencer_99, rathbun_10, rathbun_20}), and Solid-state
Imaging (SSI; \citealp{fanale_00,
  leonard_18}) instruments. Flybys with the Cassini and New
Horizons spacecraft allowed for further study of the water and hydrated mineral
composition of the surface ice \textit{via} the Visual and Infrared Mapping
Spectrometer (VIMS), Linear Etalon Imaging Spectral Array (LEISA), and LOng-Range
Reconnaissance Imager (LORRI) instruments \citep{brown_03, mccord_04,
  grundy_07}. Recently, high spatial resolution observations of the
surface with
the Jovian InfraRed Auroral Mapper (JIRAM) onboard the
\textit{Juno} spacecraft allowed for constraints on the ice grain
size, while \textit{in situ} magnetometer measurements helped to
better characterize the charged particle environment at Europa
\citep{filacchione_19, mishra_21, addison_23}.

Observations of the Galilean Satellites in support of these
missions (and in-between) have been conducted with ground- and space-based assets, 
improving our understanding of the distinct coloration, albedo
differences, and hydrate 
absorption features across
the surface. These include compositional and thermal studies utilizing the International Ultraviolet Explorer (IUE; \citealt{lane_81,
  domingue_98}), Hubble Space Telescope
(HST; \citealt{noll_95, brown_13, trumbo_20, trumbo_22}), the airborne
SOFIA observatory
\citep{de_pater_21b}, and various ground-based facilities such as the Very
Large Telescope (VLT), Infrared Telescope Facility (IRTF), and Keck
\citep{hansen_73, de_pater_89, spencer_02, spencer_06, fischer_15, ligier_16,
  fischer_17, trumbo_17a, king_22}. Recent results from the James Webb Space
Telescope (JWST) also show evidence for the endogenous origin of
previously detected surface CO$_2$, potentially sourced from the subsurface
ocean \citep{trumbo_23, villanueva_23}.
Though the variability in surface terrain and reddened trailing
hemisphere have now been well characterized, the
endo- and exogenic processes that have influenced Europa's surface
composition and evolution are currently poorly understood,
and will likely remain so until the arrival of the \textit{JUpier ICy moons
  Explorer (JUICE)} and \textit{Europa Clipper}
spacecraft in the future (which will undoubtedly provide many
additional questions of their own).

Complementary to the aforementioned observations at shorter
wavelengths are those in the radio and (sub)millimeter regime, which
probe the near-surface crust down to $\sim10$s of cm to m depths;
beyond, the deeper layers of the crust may be probed by microwave and radar observations down
to $\sim10$ km \citep{ostro_82, ostro_92, chyba_98,
  bruzzone_13}, including recent \textit{in situ} remote sensing with
the \textit{Juno} MicroWave Radiometer (MWR; \citealp{janssen_17,
  zhang_23}), and future thermal imaging, submillimeter, and radar observations from the \textit{Europa
  Clipper} and \textit{JUICE} spacecraft \citep{hartogh_13,
  phillips_14, pappalardo_17}. Millimeter wave observations at
different wavelengths permit the measurement of
thermal radiation as a function of subsurface depth, the modification
of which is governed by the thermophysical properties of the
surface. These include the
millimeter emissivity, subsurface thermal inertia, porosity, dust
fraction, and grain size, all of which inform
our understanding of how the various endo- and exogenic processes have altered the
surface, and to what extent they change the subsurface structure and
composition. Initial characterization of the subsurface properties
and thermal emission of the Galilean Satellites were made with a
number of long-wavelength facilities throughout the last half-century,
including single dish facilities such as the 2.24-m telescope on
Maunakea, the 12-m dish at Kitt Peak, the Effelsberg 100-m telescope, and the Institut de
Radioastronomie Millim{\'e}trique (IRAM) 30-m telescope
\citep{morrison_72, morrison_73, ulich_76, pauliny_toth_77, ulich_84, altenhoff_88}; the Owens
Valley Radio Observatory (OVRO) 3-element array \citep{berge_75, muhleman_91}; the 
the SubMillimeter Array (SMA) and Very
Large Array (VLA) interferometers \citep{de_pater_82, de_pater_84, muhleman_86};
and the Photodetector Array Camera and Spectrometer (PACS) onboard the Herschel space-based telescope \citep{muller_16}. Often, the
Galilean Satellites were also used for flux calibration observations for
(sub)millimeter facilities, along with Saturn's largest moon, Titan
\citep{ulich_81, moreno_07, butler_12}.

Radio and (sub)millimeter interferometric observations from modern telescopes
can spatially resolve small Solar
System bodies, such as Europa, and thus enable the measurement of
thermophysical properties as a function of location on the body, by modeling
the thermal radiation from the subsurface (cm-m depths). Utilizing the Atacama Large
Millimeter/submillimeter Array (ALMA), \citet{trumbo_18} mapped the
thermal inertia of Europa's surface using 1.3 mm (233 GHz) observations; they also
investigated the correlation of thermal anomalies observed with ALMA
with potential plume locations
\citep{trumbo_17b}. These studies revealed that a
  global thermal inertia of 95 J m$^{-2}$ K$^{-1}$ s$^{-1/2}$
  and emissivity of 0.75
  provided good fits to the ALMA observations. They found that anomalously
  cold locations in the ALMA observations, such as around Pwyll crater
  ($\sim271^\circ$W, $25^\circ$S) and a region on the leading hemisphere
  ($90^\circ$W, 23$^\circ$N), were indicative of localized, high
  thermal inertia regions or
  low emissivity; thermal inertia values ranging from
  40--300 J m$^{-2}$ K$^{-1}$ s$^{-1/2}$ or emissivities from
  0.67--0.84 were found to characterize outlying regions in the residual maps, though
  thermal anomalies were not correlated with geological or
  morphological features (excepting Pwyll). The retrieved thermal inertias are comparable to those
  derived for the surface from Galileo/PPR observations, which
provided constraints on Europa's thermal inertia from 40--150 J
m$^{-2}$ K$^{-1}$ s$^{-1/2}$, with elevated measurements in similarly anomalous
regions such as near Pwyll \citep{spencer_99, rathbun_10, rathbun_20}. Recent analyses
have also been carried out for Ganymede \citep{de_kleer_21a} and Callisto
\citep{camarca_23} using ALMA to investigate the change in porosity or
thermal inertia 
as a function of depth and correlate brightness temperatures to
geographically distinct surface regions. On Ganymede,
\citet{de_kleer_21a} found that a porosity gradient between 10--40$\%$
provided good fits to ALMA observations sounding the upper $\sim$0.5 m
of the subsurface. From ALMA Band 7 data, \citet{camarca_23} derived
a mixture of high (1200--2000 J
m$^{-2}$ K$^{-1}$ s$^{-1/2}$) and low (15--50 J
m$^{-2}$ K$^{-1}$ s$^{-1/2}$) thermal inertia components to correctly
model the thermal emission from Callisto's leading
hemisphere. Both studies found cold thermal anomalies co-located with
the locations of crater basins or complexes. Generally, these studies revealed higher thermal inertias on
the near subsurface of Ganymede and Callisto than Europa. 

Here, we present the analysis of ALMA observations of Europa at three
wavelengths (0.88, 1.25, and 3.05 mm) that probe distinct depths in Europa's subsurface, which
allows us to investigate the change in thermophysical properties with
depth and latitude, and ascertain their potential link to exogenic
sources and the evolution of Europa's ice shell. These observations
complement the recent studies of Ganymede and Callisto with ALMA, and
provide context for \textit{Juno} observations of Europa with infrared
and microwave instruments. In Section
\ref{sec:obs}, we detail the ALMA observations, reduction and imaging
procedures, followed by the radiative transfer modeling in
Section \ref{sec:rad}. A discussion of the modeling results is
presented in Sections \ref{sec:dis}, 
followed by our conclusions in Section \ref{sec:conc}.

\section{Observations} \label{sec:obs}
The ALMA Main Array is an interferometer consisting of up to 50 12-m
antennas located in the Atacama Desert, Chile. Every pair of antennas
acts as a two-element interferometer, measuring a single complex
component (often called a ``visibility'') of the Fourier transform of
the sky brightness. Together, the collection of visibilities allows
for the reconstruction of the full sky brightness in both dimensions
via image deconvolution techniques (see \citealp{thompson_01}, and
references therein). As part of ALMA
Project Code 2016.1.00691.S, the leading and trailing hemispheres of
each of the Galilean Satellites were observed in three distinct
frequency bands that probe different subsurface depths: ALMA Band 3
(97.5 GHz; 3.05 mm), Band 6 (233 GHz; 1.25 mm), and Band 7 (343.5 GHz,
0.88 mm). Europa was observed 8 times between 2016 and 2017. As the angular resolution of interferometric
observations depends on the distances between antennas in the array,
these observations were executed using different antenna configurations so as to
obtain relatively consistent resolution across all three frequency
bands. A configuration with maximum antenna separation of 6.4 km was
used for Band 3 observations to achieve comparable resolution to data from higher
frequency bands, while a
configuration with a shorter maximum antenna separation of 1.3 km was
used for Bands 6 and 7. Separate observations in
each frequency band were executed to target both the leading and
trailing hemispheres of Europa, with typical integration times of
$\sim120-300$ s; as such, longitudinal smearing over this time period
was well below the size of a resolution element. All observations were
carried out using between 40 and 45 antennas. In some cases, multiple
executions (i.e. observing integrations)
were acquired for each hemisphere in a single band, allowing for
additional longitude coverage and higher constraints on thermophysical
properties. An additional execution in ALMA Band 7 was performed, but
was set to incorrect sky coordinates, and as such was not
analyzed here. The observation parameters for each integration are detailed
in Table \ref{tab:obs}. 

\begin{deluxetable}{lcclcclrcll}
  \tablecaption{Observational Parameters}
  \tablecolumns{11}
  \tablehead{Obs. Date & Freq.$^{a}$ & $\lambda$ & Tag$\#^{b}$ &
    Ang. Diam. & Spatial Res. &
   Pos. Ang.$^c$ & Lat. & W Lon. & Corr.$^d$ \\ 
     (UTC) & (GHz) & (mm) & & (arcsec) & (arcsec) & ($^\circ$) & ($^\circ$N) &
     ($^\circ$W) & Factor \\ [-2.75ex]}
  \startdata
  %UTC date                     Freq     Wv   Hemi       Ants  Ang. Di. Beam                         PA       Lat.    Lon. W.    Corr. Fact
2017 Sep 19 17:15 & 97.5 & 3.05 & 3L & 0.684 & $0.110\times0.086$& 52.23 &-3.08 & 103.3  & 1.0 \\
2017 Sep 28 14:43$^\dag$ & & & 3T & 0.677 & $0.148\times0.085$ & 54.87 & -3.12 & 283.2 & 1.0 \\
\hline
2017 Aug 07 19:15 & 233 & 1.25 & 6L & 0.737 & $0.107\times0.082$ & 87.24 & -2.92 & 80.63 & 0.985 \\
2017 Jul 09 01:03 & & & 6T0$^e$ & 0.797 & $0.280\times0.107$ & -73.13 & -2.89 & 307.7 & 1.068 \\
2017 Jul 30 00:05 & & & 6T1 & 0.754 & $0.163\times0.078$ & -70.25& -2.91 & 270.1 & 1.028\\
2017 Aug 16 22:55 & & & 6T2 & 0.722 & $0.127\times0.079$ & -70.64 & -2.95 & 287.0 & 0.937\\
\hline
2017 Jul 06 23:56$^\dag$ & 343.5 & 0.88 & 7L & 0.803 & $0.139\times0.068$ & -68.88 &-2.90 & 100.3 & 0.956 \\
2016 Oct 25 12:28 & & & 7T$^e$ & 0.678 & $0.196\times0.151$ & 66.85 & -2.50 & 226.7 & 0.938 \\

  \enddata
  \footnotesize
  \tablecomments{$^{a}$Averaged frequency of all continuum
    windows. Frequencies correspond to ALMA Band 3 (97.5 GHz), Band 6
    (233 GHz), and Band 7 (343.5 GHz). $^{b}$Tag denoting the ALMA
  frequency band, targeted hemisphere (L =
  leading, T = trailing), and observation number; exact longitudes vary slightly
  for each execution. Hemispheres with multiple integrations are denoted
with separate labels for each individual execution. $^c$The position angle
of the synthesized ALMA beam, denoted in degrees counter-clockwise
from the positive vertical. $^d$Correction factor derived from variability of
quasars used for flux density calibrations. $^{e}$6T0: Though data from this execution were reduced and
modeled, the beam dimensions prevents the data from
yielding meaningful longitudinal information regarding Europa's surface
properties. 7T: A
second execution for the trailing hemisphere in Band 7 was not used.
$^{\dag}$Denotes observations where interloping
  satellites were present in the ALMA field - see Appendix \ref{sec:app_a}.}
  \label{tab:obs}
 \end{deluxetable}

Data from each integration were reduced using the Common Astronomy Software
Applications (CASA) package ver. 4.7 
\citep{jaeger_08} and the provided ALMA pipeline
scripts. Continuum images were produced by flagging channels with
telluric contamination and then averaging to 
channel bins of 125 or 256 MHz to reduce data volume. The resulting data were then
combined using multi-frequency synthesis imaging methods to produce a
single, high signal-to-noise ratio (SNR) broadband image of the thermal
continuum emission. Phase self-calibration was performed on each
observation to compensate for
tropospheric phase fluctuations, which improves image coherence and SNR for each observation (see the discussion in
\citealp{cornwell_99b, butler_99, brogan_18}, and ALMA Memo
620\footnote{https://library.nrao.edu/public/memos/alma/main/memo620.pdf}
by Richards et al.). Similar procedures were applied to the accompanying observations of Ganymede and
Callisto \citep{de_kleer_21a, camarca_23}.

Final image deconvolution, which removed interferometric artifacts induced by
the lack of complete antenna coverage on the sky, was performed using the CASA \texttt{tclean} task
with image sizes of $1000\times1000$ pixels of $0.01''$ size (note
that this is not the effective resolution, which is shown in Table 
\ref{tab:obs}, but simply the pixel size). Briggs weighting was applied
with a ``robust'' factor of 0, which slightly increases the weight of
data from larger antenna separations \citep{briggs_95}. The
removal of interferometric artifacts -- and thus the
  improvement of the final image quality -- for two of the ALMA
observations was facilitated by accounting for the
emission from nearby Galilean Satellites (Ganymede, Callisto) that
intervened on the relatively large ALMA Field-of-View (FOV),
introducing additional signal in the sidelobes. These procedures and
the improvements in the images are detailed in Appendix
\ref{sec:app_a}.

For each ALMA integration, the disk-averaged flux density of Europa was determined by
  fitting a disk model to the calibrated visibility data, often excluding data
  from larger antenna spacings
  (e.g. \textgreater100--200 m), which are sensitive to
    smaller scale thermal structure (such as surface variations) and not
    the total flux density. A correction to this value was made based
  on the variability of measured quasar brightnesses for each quasar
  used for each ALMA observation\footnote{ALMA Flux
  Calibrator Catalogue: https://almascience.eso.org/alma-data/calibrator-catalogue}, 
as was done for previous ALMA observations \citep{trumbo_18, de_kleer_21a}. The flux density for
each quasar was interpolated based on the measurements from the
nearest dates in the cases of Band 3 and 7 observations, where quasars
were commonly monitored. For Band 6 observations, quasar flux density
curves were derived based on the functional form detailed in
\citet{ennis_82}, using contemporaneous quasar observations in both
Bands 3 and 7 to determine the variability of flux density with
frequency. No corrections were needed for Band 3 data because the quasar
flux densities were determined on the same date as the observations. For Bands 6 and 7, we
found correction factors from $1.5-6.8\%$ were needed (Table \ref{tab:obs}). As found previously, the dependence of ALMA
on quasar observations can result in higher flux density scale calibration
uncertainties \citep{francis_20}; as a result, our quoted
uncertainties on the disk-averaged flux densities, temperatures, and emissivities are
no less than $5\%$, which are often larger than the statistical
uncertainties derived from the model fit for the flux density.

\begin{figure}
  \centering
  \includegraphics[scale=0.56]{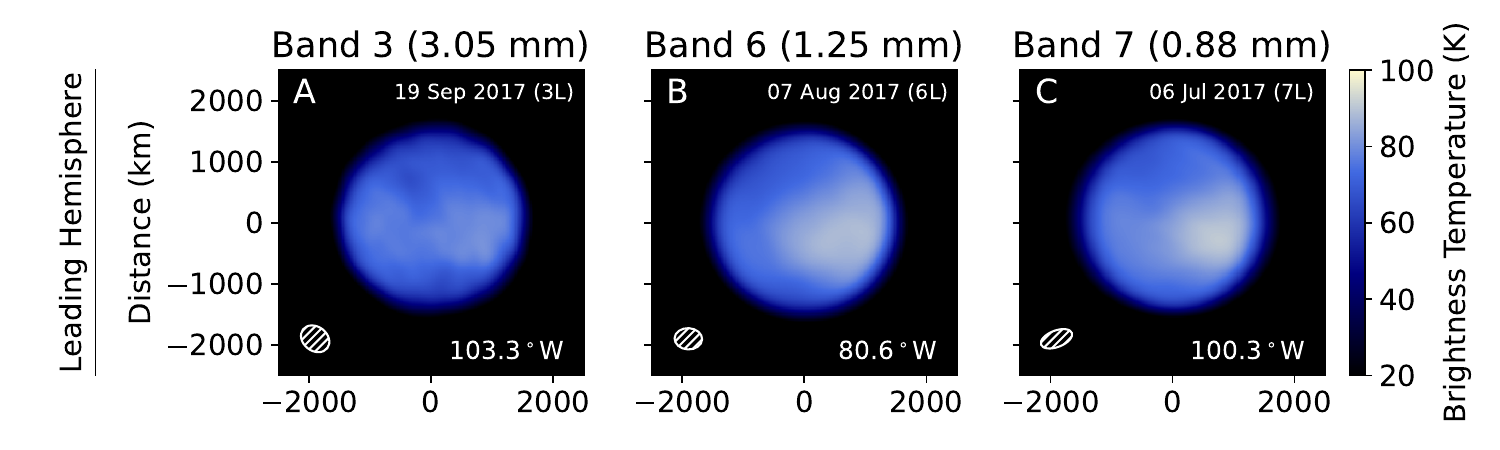}
  \includegraphics[scale=0.52]{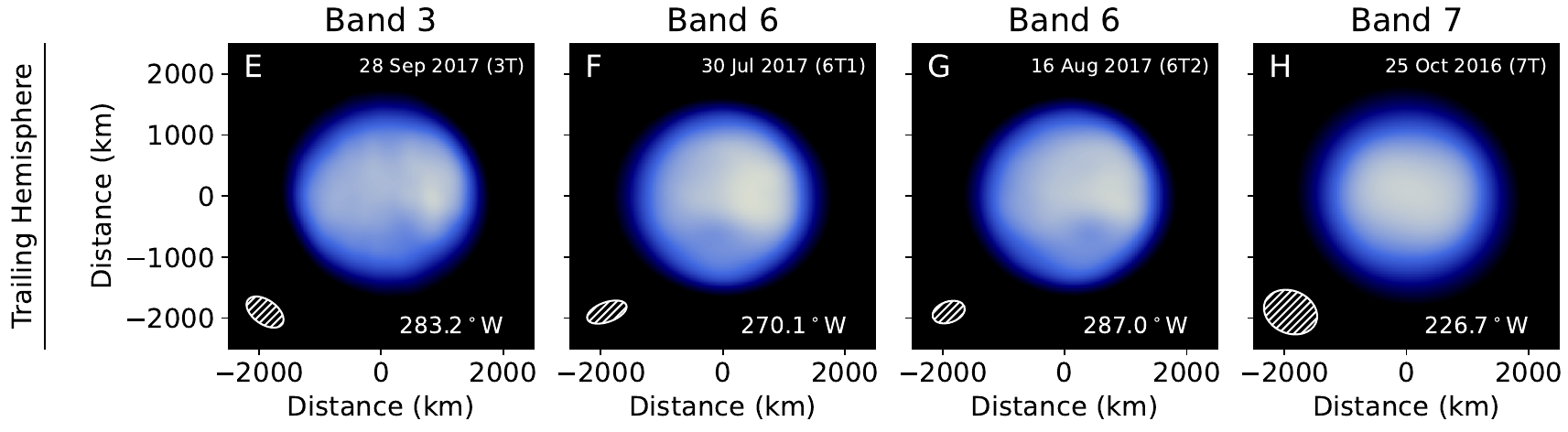}
  \caption{Brightness temperature maps of Europa's leading hemisphere
    ($\sim90^{\circ}$ W longitude; top row) and trailing hemisphere
    ($\sim270^{\circ}$ W longitude; bottom row)
    from Band 3 (3.05 mm; A, E), Band 6 (1.25 mm; B, F-H), and Band 7 (0.88 mm;
    C, G). The ALMA beam (the full-width at half-maximum of the ALMA
    point spread function) is shown as a hashed
    ellipse in the bottom left corner. All images are aligned with Europa's north pole along
      the vertical axis.}
  \label{fig:maps}
\end{figure}

After converting from flux density units (Jy) to
brightness temperature (K, the expected thermal temperature the
surface would emit if it was solely parameterized by the Planck
function; see also \citealt{de_kleer_21a} and \citealt{camarca_23}), the final emission maps were compared to radiative
transfer models generated using a variety of thermophysical properties
and global porosity or thermal inertia conditions. The Europa
continuum image maps are shown in Figure \ref{fig:maps}. The measured flux densities
and brightness temperatures are listed in Table \ref{tab:data}.

 \begin{deluxetable}{llccccc}
   \tablecaption{Derived Properties and Results}
   \tablecolumns{7}
   \tablehead{Band & W Lon. & Flux Dens.$^{a}$ & T$_B$$^{a}$ & Porosity &
     $\Gamma_{Eff.}$ & Emissivity \\
    Hemi. Tag$\#$ & ($^\circ$) & (Jy) & (K) & ($\%$) & (J m$^{-2}$ K$^{-1}$ s$^{-1/2}$) & \\ [-2.75ex]}
  \startdata
  %Obs. W lon. Flux density          Tb                 Porosity     TI eff                emiss
3L$^*$ & 103.3 & $0.18\pm0.009$ & $72.99\pm3.65$ & $50^{+20}_{-10}$ & $140^{+43}_{-70}$ & $0.79\pm0.04$ \\
3T$^*$ & 283.2 & $0.20\pm0.010$ & $85.27\pm4.26$ & $50^{+20}_{-10}$ & $140^{+43}_{-70}$ & $0.81\pm0.04$ \\
\hline
6L & 80.63 & $1.26\pm0.063$ & $76.61\pm3.83$ & $75\pm10$ &
$56^{+30}_{-24}$ & $0.86\pm0.04$\\
%6T0 & 307.71 & $1.46\pm9.93\times10^{-2}$ & $76.18\pm5.18$ & $60\pm15$ & $102\pm18$ & $0.8\pm0.01$ \\
6T1 & 270.05 & $1.37\pm0.069$ & $79.96\pm4.00$ & $40^{+15}_{-10}$ & $184^{+49}_{-64}$ & $0.84\pm0.04$ \\
6T2 & 287.02 & $1.36\pm0.086$ & $85.74\pm5.40$ & $40\pm15$ & $184^{+77}_{-64}$ & $0.83\pm0.05$ \\
\hline
7L & 100.3 & $3.34\pm0.167$ & $86.09\pm4.30$ & $60\pm15$ & $102^{+58}_{-46}$ & $0.81\pm0.04$ \\
7T & 226.7 & $2.48\pm0.154$ & $89.09\pm5.52$ & $50^{+5}_{-10}$ &
$140^{+44}_{-20}$ & $0.79\pm0.05$ \\

%Errors from adding in quadrature:
% 6L: $1.26\pm0.066$ & $76.61\pm4.00$ & $0.86\pm0.05$
% 6T1: $1.37\pm0.079$ & $79.96\pm4.58$ & $0.84\pm0.05$ 
% 6T2: $1.36\pm0.109$ & $85.74\pm6.90$ & $0.83\pm0.07$ 
% 7L: $3.34\pm0.222$ & $86.09\pm5.73$ & $0.81\pm0.05$ 
% 7T: $2.48\pm0.198$ & $89.09\pm7.10$ & $0.79\pm0.06$
  \enddata
   \footnotesize
   \tablecomments{$^{a}$Flux densities and brightness
     temperatures listed here are derived as hemispheric
     averages. $^*$Porosity and effective thermal inertia values
     derived for ALMA Band 3 observations were inferred through bounds
   as discussed in Appendix \ref{sec:app_b}.}
   \label{tab:data}
 \end{deluxetable}

\section{Thermophysical Modeling} \label{sec:rad}
The radiative transfer modeling for thermal emission of Europa follows
the procedures detailed in \citet{de_kleer_21a}, which have been used
for Ganymede, Callisto \citep{camarca_23}, and (16) Psyche
\citep{de_kleer_21b}. The model solves for thermal transport
throughout the shallow subsurface through the inclusion of thermal
conduction and radiation, solving the 1D diffusion equation with time
and depth for temperature profiles at discrete latitude and longitudes
across the observed surface. We parameterized the model for Europa
using similar fixed parameters to those for Ganymede and Callisto, where appropriate (e.g. snow and ice densities, specific
heat values). A nominal dust-to-ice fraction = 0.3
was used (similar to what was used for Ganymede by \citealp{de_kleer_21a}), though dust fraction values
between 0.1--0.5 were tested, with fairly minimal effects on the
    best-fit porosity model residuals; however, a dust fraction change
    does alter the depths at which our data are sensitive to the
    subsurface thermal emission. Similarly, models were set with an
    intermediate surface grain size of 100
    $\mu$m, and we tested models using grain sizes of 50$\mu$m--1 mm
    as relevant for Europa's
    leading and trailing hemispheres \citep{hansen_04, dalton_12, cassidy_13,
      ligier_16, 
      filacchione_19, mishra_21}. While the dispersion of small
    (\textless200$\mu$m) and large (\textgreater500 $\mu$m) grains across
    Europa likely varies with hemisphere and surface
    composition in a complex way, we find that similar porosity models
    (within the range of errors)
    provided sufficient fits to the data across the range of grain sizes. The discussion of the impact
    of grain size and other fixed parameters on the thermal
    conductivity are discussed in detail
    in \citet{de_kleer_21a}.

    The initial bolometric albedo map was generated by
    \citet{trumbo_17b} from the USGS
    Europa map\footnote{USGS controlled photomosaic map of Europa,
      2002, available at:https://pubs.usgs.gov/imap/i2757/} from Voyager and
    Galileo images, with Galileo albedo values where available
    \citep{mcewen_86} and the phase integral of 1.01 from New Horizons
    observations 
    \citep{grundy_07}; further details are provided in previous ALMA
    studies \citep{trumbo_17b, trumbo_18, de_kleer_21a,
      camarca_23}. Models were integrated over variable times steps
    (on order 1/500 Europa days) per Europa period (3.55 Earth days), 
  including periods where Europa was in eclipse behind Jupiter, for up
  to 15 Europa days until temperature profiles converged to within 0.1 K. Longitude ranges where Europa was in eclipse for each observation were retrieved from
  the JPL Horizons ephemerides
  data\footnote{https://ssd.jpl.nasa.gov/horizons/app.html$\#$/}. 
  We
  modeled thermal emission from Europa's subsurface over a range of
  10 thermal skin depths ($\sim$0.5--0.75 m for relevant temperature
  and porosity ranges). Vertical temperature profiles and
    emission angles were
    generated independently over Europa's surface in a grid
    of $5^\circ$ latitude and longitude bins; as discussed in
    \citet{de_kleer_21a}, the incorporation of Fresnel or Hapke
    refraction does not sufficiently match the limb emission due to
    surface roughness or volume scattering, and thus is not employed here. Our methodology differs from the techniques
    employed by \citet{trumbo_17b} and \citet{trumbo_18} in that
    thermal emission was integrated over depth, where as the
    aforementioned studies treated thermal emission as originating
    only from the surface (and were thus comparable to models used to
    interpret data from 
    Galileo/PPR). These properties are calculated in the model of
    \citet{de_kleer_21a}, and allow us to generate models including
    subsurface emission for a range of porosity values. Finally, an
    additional scale factor on order 10 was
  multiplied to the imaginary part of
  the index of refraction -- derived from the complex dielectric
  constant using a mixture of snow, dust, and ice properties -- such that emission from Europa's
  subsurface was properly modeled with depth and porosity. This factor
  was derived empirically through comparisons of the $\chi^2$ values
  over our porosity grid range and a range of scale factors from
  1--30, and the corresponding increase in the imaginary
    portion of the index of refraction brought our model values to between
    $1\times10^{-4}-1\times10^{-3}$, in agreement with the range of
    values measured for cold (\textless200 K) water ice at millimeter
    wavelengths (see \citealp{warren_84}, \citealp{matzler_87},
    \citealp{matzler_98}, and references therein). The
  multiplicative scale factor decreases the 
  electrical skin depth, thus increasing the absorption of
  millimeter-wave emission at the appropriate ($\sim$centimeter) depths in the model. The increased
  imaginary index 
    could be attributed to minor amounts of saline ice at depth, the effects of which
    are not well characterized at millimeter wavelengths through
    laboratory studies \citep{matzler_98}, but would change the
    effective thermal conductivity and electrical skin depth in addition to that of pure water
    ice, dust, and snow, as are currently parameterized in the thermal
    model. Without this factor, the thermal models did not provide
    good fits to the data, and the retrieved
  best-fit porosity values were low (e.g. 10--20$\%$),
  corresponding to thermal
  inertia values approaching that of solid ice.

The thermophysical model of \citet{de_kleer_21a} can be run in two
modes: in the ``thermal inertia'' mode, the thermal inertia and
electrical properties of the material are fixed, such that the thermal
properties do not change with depth, time, or temperature. In this
mode, the model is similar to thermophysical models typically used to
interpret IR data, except that emission is integrated through the
subsurface as is necessary for interpreting radio and millimeter-wave
data. In the second, ``porosity'' mode, the subsurface porosity is the
primary free parameter and controls both the thermal and electrical
properties in a self-consistent way. All material properties (and thus
thermal inertia) vary with temperature and density, and hence with
depth and time, such that we can only report an ``effective thermal
inertia'' ($\Gamma_{eff}$) for these models. We ran models over a grid
of porosity values from 10--90$\%$, as well as single thermal inertia
models ranging from 20--1000 J m$^{-2}$ K$^{-1}$ s$^{-1/2}$, covering
values that have been observed throughout the Solar System icy bodies
\citep{ferrari_18}. The resulting porosity or thermal inertia models
were then subtracted from the data, and comparative $\chi^2$ values
determined for the residual fits resulting in the best-fit hemispheric
thermophysical properties. We found that thermal inertia models were
able to produce adequate fits in addition to those using porosity, and
compared well to the derived effective thermal inertia, defined as:

\begin{equation} \label{eq:TI}
  \Gamma_{eff} = \sqrt{k_{eff}(p, R, T_{eff})\rho_{eff}(p)c_p(T_{eff})}
\end{equation}
Here, $k_{eff}$ is the effective thermal conductivity of the ice as
a function of porosity ($p$), grain size ($R$), and effective
temperature ($T_{eff}$); see Section 3.3 of \citet{de_kleer_21a} for
the derivation of $k_{eff}$. $\rho_{eff}$, the effective density, is a function of the surface
density ($\rho_s$) and porosity: $\rho_{eff}(p) = \rho_s \times (1-p)$. Finally,
$c_p(T_{eff})$ is the effective heat capacity. However, the porosity
models incorporate the change in thermal emission as a function of
depth throughout the subsurface, and are thus more physically
realistic; further, we tested porosity for each ALMA frequency band
and hemisphere independently to determine if a compaction length could
be readily derived from the resulting porosity values. This is
discussed further in Section \ref{sec:dis}

  \section{Results $\&$ Discussion} \label{sec:dis}
\begin{figure}
  \centering
  \includegraphics[scale=0.6]{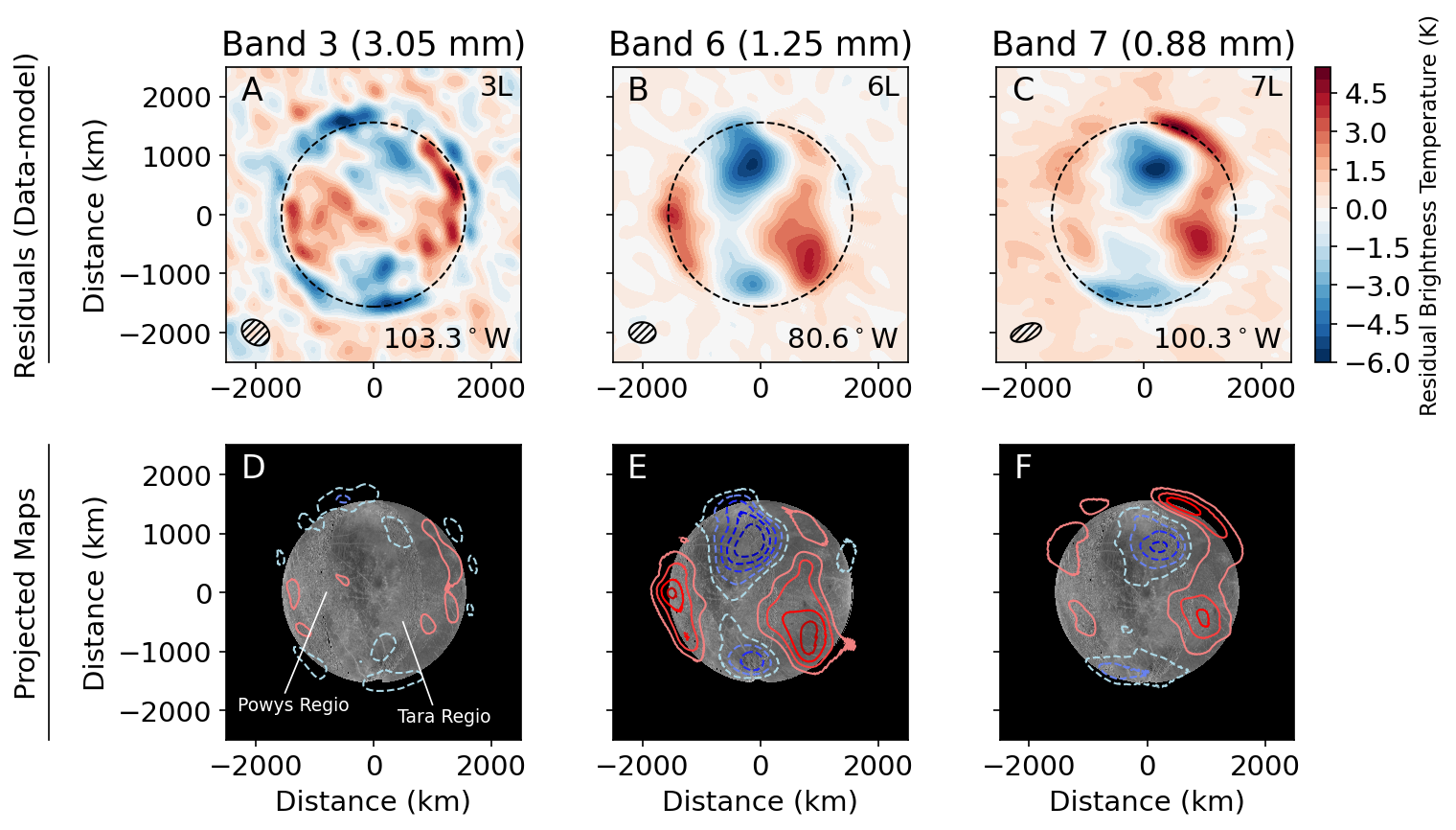}
  \caption{\textit{Top:} Residuals (data-model) for single,
    hemispheric best-fit porosity values for Europa's leading
    hemisphere ($\sim90^{\circ}$ W longitude) from Band 3 (97.5 GHz/3.05 mm; A), Band 6 (233 GHz/1.25 mm; B), and Band 7 (343.5 GHz/0.88 mm;
    C). All images are aligned with Europa's north pole along
      the vertical axis. \textit{Bottom:}
    Residual contours are plotted on projected image maps of Europa's
    surface from the USGS Voyager and Galileo SSI composite
    map. Positive temperature
    contours are shown as redscale, solid lines; negative contours are
    in bluescale, dashed lines. Contour
      levels increase in increments of $3\sigma$ 
      (RMS noise varies between observation, on the order of 0.1-1
      K). The approximate locations of leading hemisphere regiones
        are denoted for reference in panel D.}
  \label{fig:res_L}
\end{figure}

\begin{figure}
  \centering
  \includegraphics[scale=0.45]{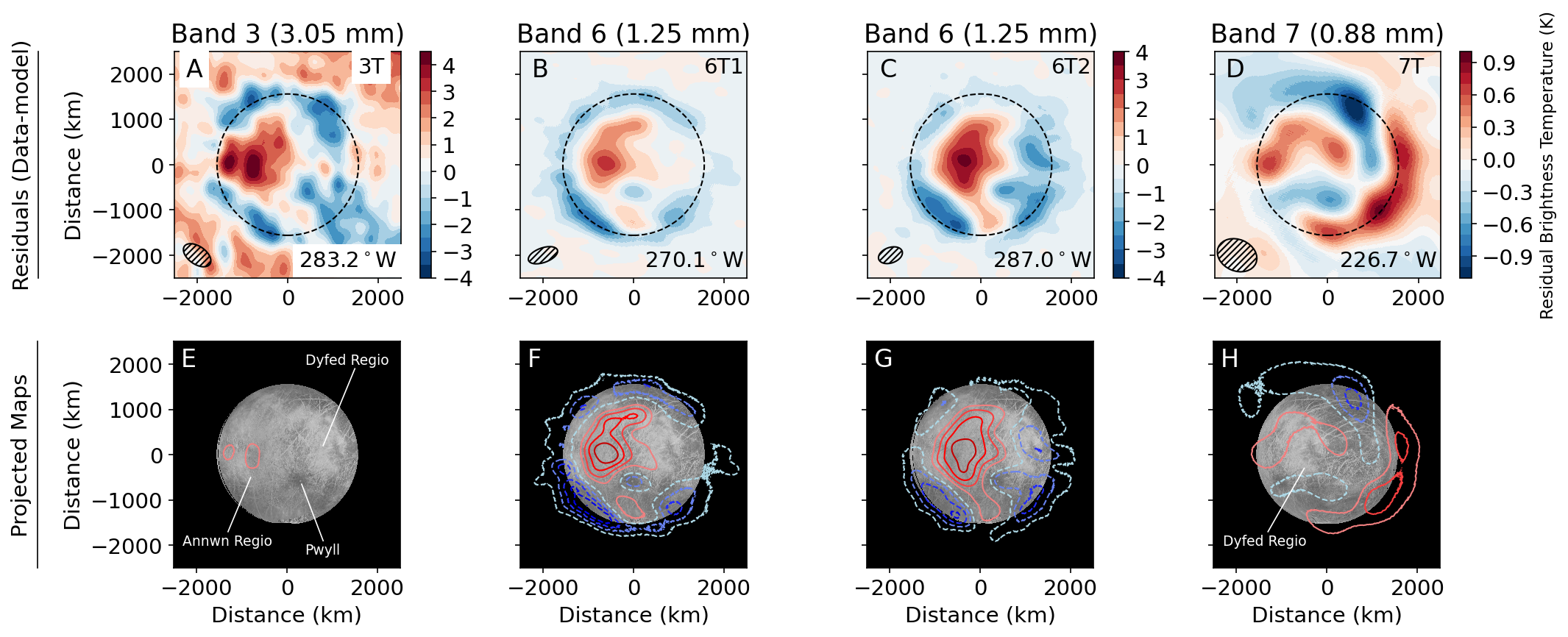}
  \caption{\textit{Top:} Residuals (data-model) for single,
    hemispheric best-fit porosity values for Europa's trailing hemisphere
    ($\sim270^{\circ}$ W longitude) from Band 3 (A),
    Band 6 (B-C), and Band 7 (D). Note the different colorbars between the three ALMA bands. All images are aligned with Europa's north pole along
      the vertical axis. \textit{Bottom:}
    Residual contours are plotted on projected image maps of Europa's
    surface from the USGS as in Figure \ref{fig:res_L}. Contour
      levels increase in increments of $3\sigma$, e.g. $3\times$ the
      image RMS noise. The approximate locations of regiones
        and Pwyll crater are denoted for reference in panels E and H.}
  \label{fig:res_T}
  \end{figure}

The residuals from the best-fit models are
shown in Figure \ref{fig:res_L} and \ref{fig:res_T} for the leading
and trailing hemispheres, respectively, with projections of Europa's
surface terrain for reference\footnote{Projection maps of Europa are able to be generated
      here:https://astrocloud.wr.usgs.gov/index.php}. The best-fit values for porosity models
and their corresponding emissivity values, as well as the converted
$\Gamma_{eff}$ for each porosity, are given in Table
\ref{tab:data}. 

Unlike in the work of \citet{camarca_23} regarding Callisto, we were
able to achieve a good fit to the data (i.e. a single
  parameter set produced a global
  $\chi^2$ minimum and significantly smaller residuals than other models) using only a single porosity or
thermal inertia value for each ALMA image. The
temperature residuals from the best-fit model were on the order of, or slightly lower than,
those found by \citet{trumbo_18} for Europa in ALMA Band 6 (1.25 mm). Though
\citet{de_kleer_21a} tested a simultaneous fit to all
Ganymede longitudes to retrieve porosity values,
we attempted to fit individual images to investigate potential
differences between the leading and trailing hemispheres. We report a distinct difference
between the best-fit properties for each
imaged hemisphere. The images targeting the leading
hemisphere yield porosities that decrease from
  $\sim70\%$ to 50$\%$ from observations at $\lambda=0.88$
  and 1.25 mm to $\lambda=3.05$ mm, while on the trailing hemisphere, slightly lower
  porosity values of 40--50$\%$ were retrieved. 
Using Equation \ref{eq:TI}, the above porosities represent a range of
effective thermal inertiae from 56--184 J
  m$^{-2}$ K$^{-1}$ s$^{-1/2}$. The upper and lower bounds on porosity -- and as a result, the
retrieved effective thermal inertia and emissivity ranges -- were
determined through $\chi^2$ statistics as in other works \citep{hanus_15,
  de_kleer_21b, cambioni_22}, defining the representative range in which similar
models provide sufficient solutions to the data with reference to the
minimum $\chi^2$ model. These final results are summarized in Table
\ref{tab:data}. The emissivity values reported here are those of the
material integrated over the viewing pathlength, as opposed to from the surface
emission as determined through IR observations. 
   
We note that porosity and thermal inertia fits for data from ALMA Band
3 ($\lambda=3.05$ mm), in both
hemispheres, showed very similar residual patterns; determining the
best-fit parameters from $\chi^2$-minimization alone was not
sufficient (i.e. there was not a clear, global $\chi^2$ minimum) due to the similar residual patterns and relatively
low SNR. As a result, the porosity values for observations at $\lambda=3.05$ mm are
inferred through upper and lower bounds determined by the 
best-fit porosities from the $\lambda=0.88$ and 1.25 mm data (under the
assumption that porosity does not increase with depth), and the depth at which the electrical and
thermal skin depths are equal, respectively. The latter bound is set
due to the lack of significant thermal anomaly features observed in the $\lambda=3.05$ mm
data compared to those in $\lambda=0.88$ and 1.25 mm (see Figure
\ref{fig:res_L}, panels A, D, 
and Figure \ref{fig:res_T}, panels A, E), and the small effects that
varying thermal inertia and porosity models have on the residual fits;
together, these properties indicate that the ALMA observations at
$\lambda=3.05$ mm 
are sensitive to subsurface layers below a thermal skin depth, where
diurnal temperature variations are significantly diminished. Further discussion is provided in Appendix
\ref{sec:app_b}. 

\subsection{Derived Thermophysical properties} \label{sec:tp_prop}
\begin{figure}
  \centering
  \includegraphics[scale=0.45]{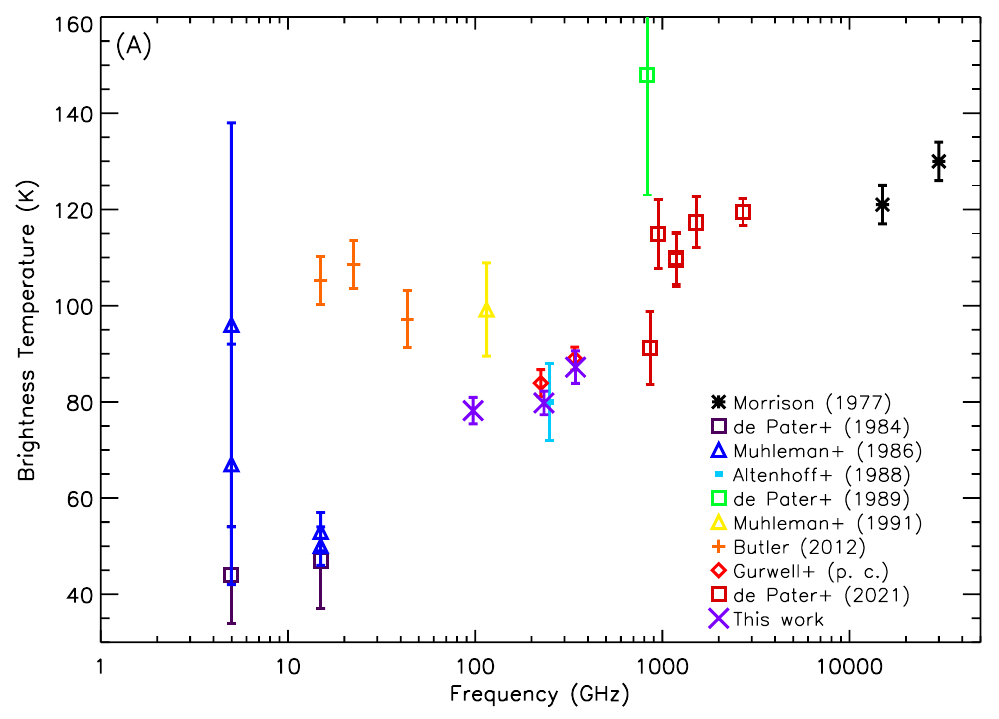}
  \includegraphics[scale=0.45]{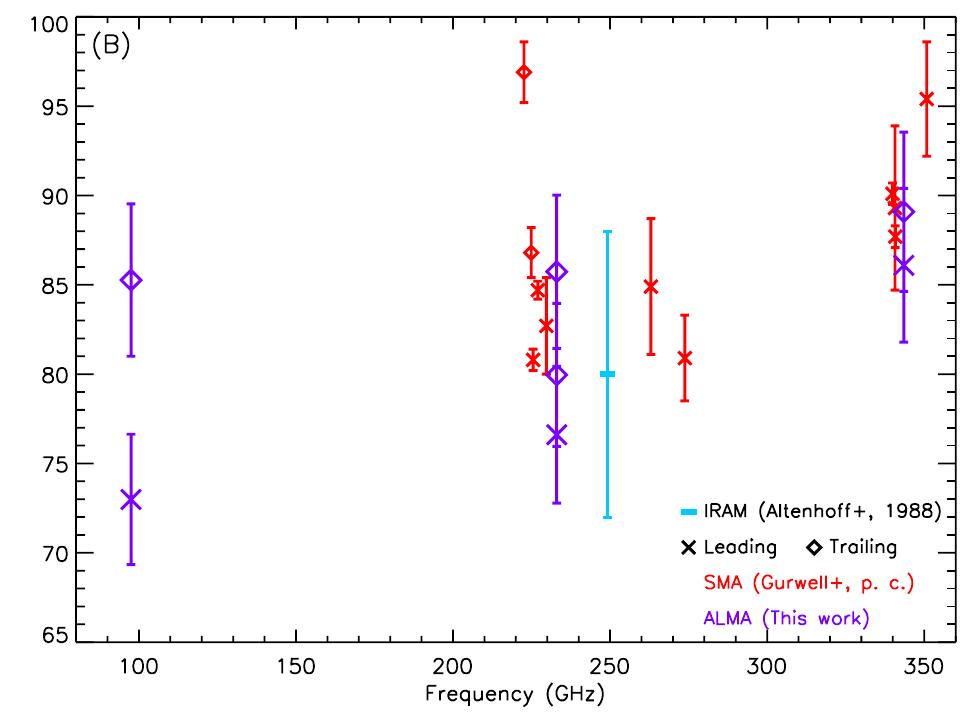}
  \caption{(A) Comparison of Europa's disk-averaged brightness temperatures as a
    function of frequency from ALMA (weighted averages from this work; purple), other
    radio/(sub)mm facilities including the VLA
    \citep{de_pater_84, muhleman_86, butler_12}, 
    the IRAM 30-m dish \citep{altenhoff_88}, the SMA
    (Gurwell et al., private communication), OVRO \citep{muhleman_91},
    and infrared measurements using SOFIA
    \citep{de_pater_21b} and the NASA IRTF
    \citep{de_pater_89}. Predictions of the maximum surface
    temperatures at visible wavelengths from \citet{morrison_77}
      are shown in black. (B) Brightness temperatures as a function of frequency
  in the (sub)millimeter wavelength regime from both the leading
  (crosses) and trailing (diamonds) hemispheres as measured by ALMA
  (this work; purple) and the SMA (Gurwell et al., private
  communication; red). Data from the IRAM 30-m
  telescope are also shown (\citealp{altenhoff_88}; blue).}
\label{fig:da_tb}
\end{figure}

The weighted mean of our derived temperature and
  thermophysical properties are listed in Table \ref{tab:data_av}. Our
  mean, disk-averaged brightness temperatures are
compared to previous measurements of Europa 
at thermal wavelengths in Figure \ref{fig:da_tb} (panel A). Measurements from
each hemisphere are compared in Figure \ref{fig:da_tb} (panel B). We observe an increasing
divergence in hemispheric brightness temperature with wavelength
(decreasing frequency in Figure \ref{fig:da_tb}, panel B), though this trend is
only significant at lower frequencies (Band 3; $\lambda=3.05$ mm).
Temperatures derived from the ALMA $\lambda=0.88$ and 1.25 mm
observations are in good agreement with previous measurements from the IRAM 30-m telescope
\citep{altenhoff_88} and SMA data acquired between 2008 and 2022
(Gurwell et al., private communication) at similar wavelengths. The
SMA measurements show a similar hemispheric disparity to our ALMA Band 6
observations, and corroborate the decrease in brightness temperature with wavelength (Gurwell et al.,
private communication). It is unclear what the exact central longitude of Europa was during
the observations of \citet{altenhoff_88}, but it appears to be of Europa's
leading to anti-Jovian hemisphere ($\sim90-180^\circ$W), and is
similar to both ALMA and SMA measurements of the leading hemisphere. Our measurements at
$\lambda=3.05$ mm are lower than those
found by \citet{muhleman_91} with the OVRO, although the value shown
in Figure \ref{fig:da_tb} (panel A) from that study corresponds to the trailing hemisphere of
Europa, which is more in line with our measurements than for the
leading hemisphere. Further observations with ALMA Band 4 and 5
(125--211 GHz; 1.4--2.4 mm) and the VLA could help determine if
the observed hemispheric disparity is consistent with the
$\lambda=3.05$ mm
observations and persists down to $\sim$ m depths. VLA observations at
additional Europa longitudes would make for interesting comparisons with
  previous analyses by \citet{de_pater_84}, \citet{butler_12}, and
  \citet{muhleman_86}.
  
 \begin{deluxetable}{lccccc}
   \tablecaption{Globally Averaged Subsurface Properties}
   \tablecolumns{6}
   \tablehead{ALMA & Depth & T$_B$ & Porosity &
     $\Gamma_{Eff.}$ & Emissivity \\
    Band & (cm) & (K) & ($\%$) & (J m$^{-2}$ K$^{-1}$ s$^{-1/2}$) & \\ [-2.75ex]}
  \startdata
  %Band    z     Tb                 Porosity     TI eff                emiss
3 (3.05 mm) & $\sim0.5-1$ & $78.19\pm2.77$ & $50^{+20}_{-10}$ & $140^{+43}_{-70}$ & $0.80\pm0.03$ \\
6 (1.25 mm) & $\sim1.5-3$ & $79.78\pm2.46$ & $64\pm8$ & $76\pm25$ & $0.85\pm0.02$ \\
7 (0.87 mm) & $\sim10-20$ & $87.22\pm3.39$ & $52\pm7$ & $130\pm27$ & $0.80\pm0.04$\\
  \enddata
   \footnotesize
   \tablecomments{Properties listed are the weighted average of those
     detailed in Table \ref{tab:data}.}
   \label{tab:data_av}
 \end{deluxetable}

In contrast to the brightness temperatures derived for the other
Galilean Satellites (see, e.g. \citealp{de_kleer_21a},
\citealp{de_pater_21b}, and \citealp{camarca_23} for recent work and
literature comparisons) and Pluto \citep{lellouch_16}, those measured at Europa do not appear to be
monotonically increasing as a function of frequency (Figure
\ref{fig:da_tb}, panel A). However, new measurements at frequencies \textless10
GHz are needed to confirm the discrepancies and large uncertainties found in early VLA
observations \citep{de_pater_84, muhleman_86}. As noted in previous
works (e.g. \citealp{de_kleer_21a}), the decrease in brightness temperature measured
with descending frequency across the ALMA wavelength range is
indicative of the colder temperatures at depth where both the
thermal inertia increases and, in the case of our ALMA Band 3
measurements, the emission is sourced from below the thermal skin depth.

We find the weighted averages of our leading and trailing hemisphere
porosity values to range between $50-64\%$, corresponding to
$\Gamma_{eff}=76-140$ J m$^{-2}$ K$^{-1}$ s$^{-1/2}$ (Table
\ref{tab:data_av}).
Our emissivity values are towards the
higher end of the range found by \citet{trumbo_18} for ALMA Band 6
(0.67--0.84), while our Band 6 value (both globally and, in
particular, on the leading hemisphere) is more towards that derived by the Voyager Infrared
Interferometer Spectrometer and Radiometer (IRIS) instrument for
the surface (0.9; \citealp{spencer_87}). It is reasonable, however,
that the measured (sub)millimeter emissivity is lower than those
derived from infrared measurements, as has been found with other
objects \citep{lellouch_16, lellouch_17, brown_17,
  de_kleer_21a}. Comparisons of our globally averaged thermal
  inertia values to previous measurements of Europa and the other
  Galilean Satellites from the surface to $\sim10$s of cm are listed in Table \ref{tab:lit}. A high porosity, low
thermal inertia surface for Europa was initially 
inferred from ground-based eclipse observations at 10 $\mu$m
\citep{hansen_73}, and a range of $\Gamma$=40--150 J m$^{-2}$ K$^{-1}$
s$^{-1/2}$ was found from the Galileo/PPR data across the
surface \citep{spencer_99, rathbun_10, rathbun_14,
  rathbun_20}. The ALMA Band 6 observations analyzed by
\citet{trumbo_18} resulted in a global average thermal inertia of 95 J
m$^{-2}$ K$^{-1}$ s$^{-1/2}$ and a typical range of $\sim40-300$ J
m$^{-2}$ K$^{-1}$ s$^{-1/2}$, when considered with their best-fit
emissivity value of 0.75. These values fall within the range of the
Galileo/PPR measurements, and our measured ALMA Band 6 average
is similar to their best-fit, global thermal inertia value, despite differences
between the models with regards to the treatment of subsurface
emission. Our Band
7 average,
though larger, still falls within the range of previously measured values, as
well as those found by \citet{trumbo_18} in various portions of the
surface at slightly lower depths. The derived thermal inertia values from the ALMA observations
fall closer to the higher thermal inertia component of the 2-component
model of \citet{spencer_87} using Voyager observations.

\begin{deluxetable}{llll}
   \tablecaption{Measured Thermal Inertiae of the Galilean Satellites}
   \tablecolumns{4}
   \tablehead{Object & $\Gamma^{\dag}$ & Facility/ & Ref. \\
   & (J m$^{-2}$ K$^{-1}$ s$^{-1/2}$) & Instrument & \\ [-2.75ex]}
  \startdata
  %Object Thermal In. Facility Ref.
Europa & $14\pm5$, \textgreater300 (2C) & Hale Observatory & \citet{hansen_73} \\
 & \textless40$^{a}$ & Maunakea 2.24-m & \citet{morrison_73} \\
 & $50\pm5$ & Voyager/IRIS & \citet{spencer_87} \\
 & $16\pm2$, $300\pm200$ (2C)  & & \\
 & 70 & Galileo/PPR & \citet{spencer_99} \\
 & $40-150$  & & \citet{rathbun_10} \\
 & 95 & ALMA & \citet{trumbo_18} \\
 & $40-300$ & & \\
 & $87, 105^{b}$ & Galileo/PPR & \citet{rathbun_20} \\        
 & $140^{+43}_{-70}$ & ALMA (Band 3) & this work \\
 & $76\pm25$ & ALMA (Band 6) & \\
 & $130\pm27$ & ALMA (Band 7) & \\   
\hline
Ganymede & $14\pm2$, \textgreater300 (2C) & Maunakea 2.24-m & \citet{morrison_73} \\
 & $70\pm20$ & Voyager/IRIS & \citet{spencer_87} \\
 & $22\pm2$, $500\pm100$ (2C) & & \\
 & $16\pm6$, $1000\pm500$ (2C) & & \\
 & $750^{+200}_{-350}$ & ALMA (Band 3) & \citet{de_kleer_21a} \\
 & $350^{+350}_{-250}$ & ALMA (Band 6) & \\
 & $450^{+300}_{-250}$ & ALMA (Band 7) & \\ 
\hline
Callisto & $10\pm1$, \textgreater300 (2C) & Maunakea 2.24-m & \citet{morrison_73} \\ 
 & $50\pm10$ & Voyager/IRIS & \citet{spencer_87} \\
 & $15\pm2$, $300\pm200$ (2C) & & \\
 & $600-1800$ & ALMA (Band 7) & \citet{camarca_23} \\
 & $15-50$, $1200-2000$ (2C) & & \\
\hline
Io & $38\pm3$, \textgreater300 (2C) & Hale Observatory & \citet{hansen_73} \\
 & $13\pm4$, \textgreater300 (2C) & Maunakea 2.24-m & \citet{morrison_73} \\ 
 & 56, 5${^c}$ & IRTF & \citet{sinton_88} \\
 & 25, 100${^c}$ & HST & \citet{kerton_96} \\
 & 70 & Galileo/PPR & \citet{rathbun_04} \\
 & 40, 100${^c}$ & Galileo/PPR & \\
 & $20\pm10, 200\pm50{^c}$ & HST, Galileo/PPR & \citet{walker_12} \\
 & 50$^d$ & Gemini/TEXES & \citet{tsang_16, de_pater_20} \\
 & 320$^d$ & ALMA & \citet{de_pater_20} \\
 
  \enddata
   \footnotesize
   \tablecomments{$^{\dag}$Best-fit values or ranges across the
     surface are listed, depending on the data analyzed. Measurements
     represent the thermal inertia of Europa's surface in some
     instances (e.g. infrared measurements) down to $\sim10$s of cm
     (e.g. ALMA Band 3). Models using
     2 thermal inertia components are denoted as
     `2C', and include values for both model components. $^a$The value for
     Europa in \citet{morrison_73} was esimated only. $^{b}$Values listed refer to proposed plume locations on
     Europa. $^c$Values listed for Io correspond to frost and
     non-frost-covered surfaces. $^d$Values derived for eclipse
     cooling of Io based on the Texas Echelon Cross Echelle Spectrograph (TEXES) instrument on the Gemini
     telescope \citep{tsang_16} and ALMA observations \citep{de_pater_20}.}
   \label{tab:lit}
 \end{deluxetable}

Our retrieved values are consistent with previous studies indicating
that Europa's surface is covered in young, refractory regolith that
may extend down to
\textgreater m depths, as probed
by radar \citep{moore_09}. From the range of previously derived
thermal inertia values at $\sim$millimeter depths (Table
\ref{tab:lit}), Europa likely has a more porous surface than
what we find for the upper $\sim10$s of cm, which changes to a less porous, higher thermal inertia
subsurface within $\sim10$s of mm. The relatively low spread of our porosity
results -- both in average and hemispheric quantities -- indicates
that Europa's subsurface porosity does not change significantly over
the top $\sim1-20$ cm of regolith. However, the derivation of a
compaction length scale (as was done for Ganymede by
\citealt{de_kleer_21a}) may be possible with future ALMA
studies at other frequencies. We find that the $\Gamma_{eff}$ values
are lower than those found for Ganymede and Callisto in the near subsurface
\citep{de_kleer_21a, camarca_23}, and like Ganymede in being much lower
than solid ice ($\Gamma$ = 2000 J m$^{-2}$ K$^{-1}$
s$^{-1/2}$). Though thermal inertia values of the Galilean Satellite surfaces
are generally larger than those of the icy Saturnian
  satellites as found by the Cassini Composite Infrared Spectrometer
  (CIRS; \citealp{howett_10, howett_14, howett_16, ferrari_18}), Cassini microwave observations of Iapetus and Rhea reveal
  elevated thermal inertiae ($\Gamma$\textgreater100) at depths of a
  few meters \citep{le_gall_14b, bonnefoy_20, le_gall_23}. The thermal
  inertiae derived for both the Galilean and Saturnian satellites at depth are larger still
  than those found for Pluto, Charon, Centaurs, Trans-Neptunian
  Objects, and main-belt asteroids using Herschel, ALMA, and the VLA, where typically $\Gamma$\textless30 J m$^{-2}$ K$^{-1}$
s$^{-1/2}$ or even of order unity \citep{keihm_13, lellouch_16, lellouch_17}.

\subsection{Hemispheric Dichotomies and Thermal Anomalies} \label{sec:hem_dis}
Fitting for the properties of each ALMA integration independently
allowed us to investigate the previously observed 
differences between Europa's leading and trailing hemispheres at
$\sim$cm depths. Although these differences are rendered somewhat
minor due to the large uncertainties, we
indeed find differences in the measured brightness temperature and 
best-fit porosity between each hemisphere across ALMA
frequency bands (Table \ref{tab:data}, Figure
\ref{fig:da_tb}, panel B); the conversion from porosity to Europa's effective
thermal inertia (Equation \ref{eq:TI}) makes this dichotomy more
  apparent. We generally find that Europa's trailing subsurface is warmer and
  less porous -- or with elevated $\Gamma_{eff}$ -- compared to the
  leading hemisphere. While our hemispheric porosity and thermal inertia models do not provide perfect
fits to the data, the single value porosity models
  yield residuals often \textless5 K. Localized
  anomalous temperature features correspond to areas of high porosity or emissivity (positive residuals), or less porous,
  less emissive, elevated thermal inertia
  surfaces (negative residuals). Overall, we find higher magnitude
  negative thermal anomalies than positive ones, particularly on the
  leading hemisphere; as a result, there exists a range of porosities and thermal
  inertiae corresponding to the largest thermal features that
  are not well described by the global average values presented in Table \ref{tab:data_av}.

While the best-fit $\Gamma_{eff}$ values are lower on the leading hemisphere,
the larger magnitude negative residuals may indicate high
thermal inertia regions at the mid-latitudes (compare negative
residuals in Figure \ref{fig:res_L} and \ref{fig:res_T}). The Band 6
and 7 trailing hemisphere observations (6T1, 6T2, and 7T) are generally better fit by a
single porosity or thermal inertia value, with the largest residuals
being towards the limb (those off-disk are likely artifacts induced
through minute differences in model and data positioning) and at equatorial latitudes towards the
center of the trailing hemisphere. In particular, the lowest magnitude
residuals are found in the anti-Jovian swath
mapped with the Band 7 observations (7T; Figure \ref{fig:res_T},
panels D, H). Here, the model provides a fit to the data to within $\pm1$ K, indicating
that a near-surface ($\sim1$ cm depth) porosity of 50$\%$ -- or an
effective thermal inertia of 140 J m$^{-2}$ K$^{-1}$ s$^{-1/2}$ -- may
be sufficient to describe the large banded and ridged plains that
cover the surface from $\sim150-240^\circ$W \citep{leonard_17}, or
that the processes that generate inhomogeneous porosity surfaces on the
other hemispheres are not as efficient here. 

Variations in our thermal residuals could be due to emissivity or
thermal inertia variations across the surface -- the former an
indication of physical (sub)surface properties (e.g. surface
roughness, subsurface
  dielectric properties, grain sizes) that were not correctly accounted for in our
model of Europa's regolith. Rough or irregular
  terrain would elevate surface temperatures; this, along with volume
  scattering, are facets to be added to the model in the future. \citet{trumbo_18} found residuals
across the disk between $\sim10$ and -8 K, which could be accounted for by varying the
emissivity by $\pm10\%$ of their derived best-fit
  value of 0.75; alternatively, the anomalies could be inferred as
thermal inertia variations ranging from 40--300 J
m$^{-2}$ K$^{-1}$ s$^{-1/2}$ or more. In our case, the 
largest magnitude residuals are smaller (+5 to -6 K), but deviations from the
best-fit models remain. These may similarly be expressed as variations in
emissivity from $\sim0.75-0.9$ and thermal inertia values
  \textless50 (warm residuals) or \textgreater200-300 J
m$^{-2}$ K$^{-1}$ s$^{-1/2}$ for the coldest residuals. As in
\citet{de_kleer_21a}, higher thermal inertia models produce
diminishing improvements in model comparisons, preventing the highest
negative residuals (those on the leading hemisphere at mid-latitudes)
from being well quantified. Positive residuals are likely elevated
porosity (or low thermal inertia) surfaces, indicating that localized regions on
both the subJovian leading and trailing hemispheres are highly
porous from the surface down to $\sim1-3$ cm.

\begin{figure}
    \centering
    \includegraphics[scale=0.62]{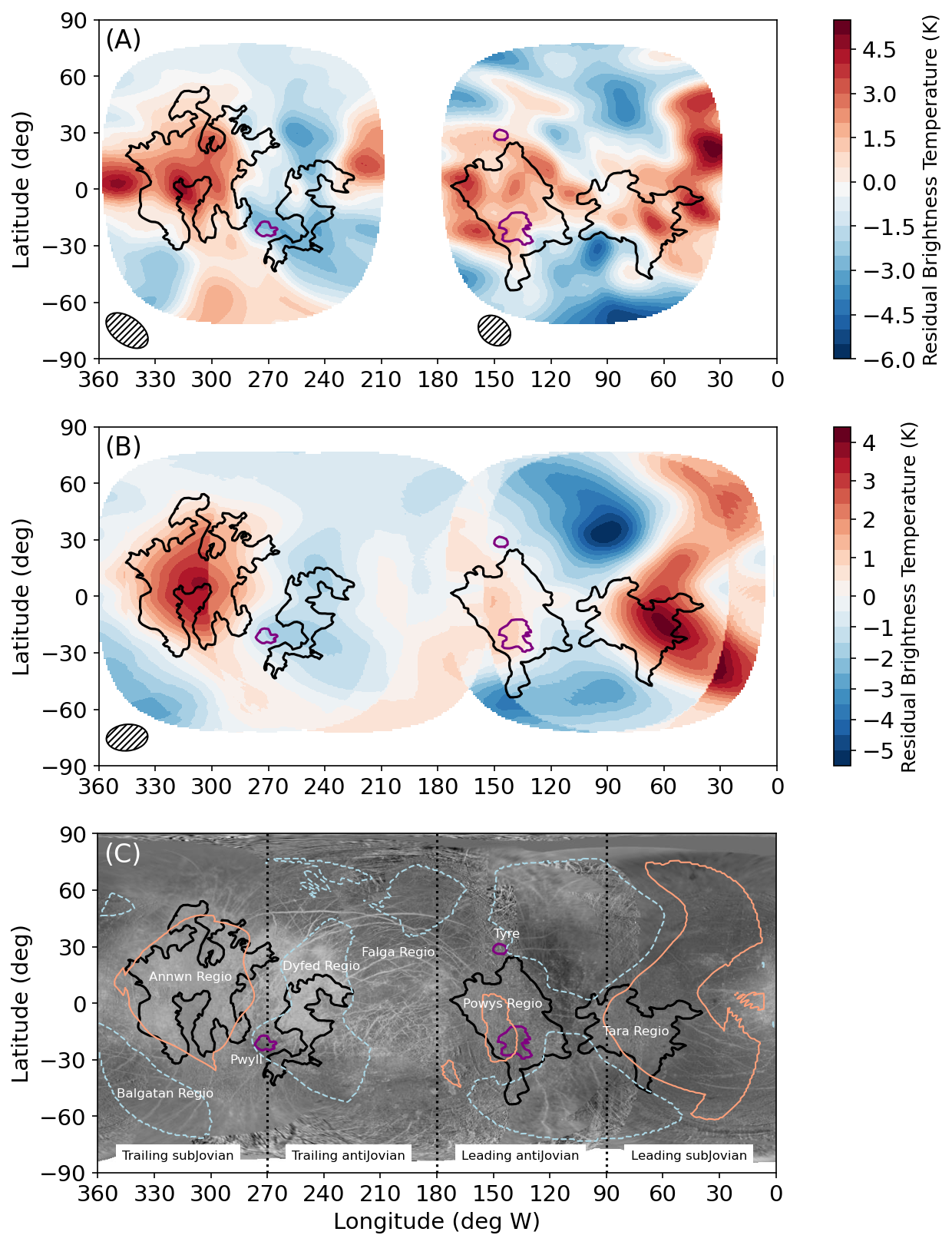}
    \caption{(A) ALMA Band 3 (3.05 mm) residuals from $50\%$ porosity
      models projected into
      cylindrical latitude and longitude coordinates, and only
      including data for emission angles less than 75$^\circ$. The
      synthesized ALMA beam for each observation (without projecting to
      cylindrical coordinates) is shown below both
      leading and trailing hemisphere projections as a hashed ellipse.
      (B) Combined residuals from Band 6 (1.25 mm) and
      Band 7 (0.88 mm) models of the global average
      values of Europa's leading and trailing
      hemisphere ($64\%$ in Band 6, $52\%$ in Band 7) projected in
      cylindrical latitude and longitude coordinates. The
      representative average ALMA beam over all Band 6 and 7 observations is shown as
      the hashed ellipse for comparison. (C)
      Averaged positive (light red, solid contours) and negative
      (light blue,
      dashed contours) residuals from (B) with magnitudes 
      \textgreater$3\times$ the average RMS ($\sim0.8$ K) overlaid onto a deprojected mosaic
      of Europa's surface from Galileo SSI and Voyager 
      images. Europa's surface quadrants are demarcated by dotted
      lines \citep{doggett_09}. Relevant geographic features are
      labeled and approximate
    outlines defined by \citet{leonard_17} are shown in all panels: black contours show Europa's chaos
    regions, and purple contours show the locations of
      the ringed terrain and ejecta blankets surrounding Pwyll, Tyre,
      and Taliesin craters. Artifacts
    exist in both the colormap in (B) and contours in (C) due to the
    combination of data from multiple executions in both ALMA
    bands.}
    \label{fig:res_map}
  \end{figure}

To better facilitate the comparison of thermal anomalies to known geological and
compositional terrain, we projected ALMA residual maps into
cylindrical coordinates shown in Figure
\ref{fig:res_map}. The models here are generated for the
  global average values listed in Table \ref{tab:data_av}, so that
  anomalies represent deviations from the global average as opposed to
  hemispheric best-fits.
Latitudes
  corresponding to large (\textgreater75$^\circ$) emission angles were
  excluded due to edge artifacts. 
  Though the depths probed by the ALMA Band 6 and 7 ($\lambda=1.25$ and
  0.88 mm) measurements are
  different, the residual patterns in Figures \ref{fig:res_L} and
  \ref{fig:res_T} are largely similar between the two where projected
  longitude ranges overlap. We combined
  Band 6 and 7 observations (including overlapping regions through
  averaged measurements)
  into a single residual map, which comprises most surface longitudes
  (Figure \ref{fig:res_map}, panel B). As
the Band 3 residuals are not as statistically significant as those exhibited by
the Band 6 and 7 data, they were not
included in this average, but are shown for comparison in Figure
\ref{fig:res_map} (panel A). Some artifacts occur where the Band 6 and 7 residual maps overlap,
and minor discrepancies between residual magnitudes
exist, particularly on the leading hemisphere. Additionally, as these measurements probe different depths in
the subsurface and different portions of Europa's day, this map is
used only for comparative purposes. However, this combined
distribution reveals the significant positive thermal distributions
across both the subJovian leading ($\sim30-90^\circ$W)
and trailing ($\sim270-330^\circ$W) hemispheres, and the negative
residuals present on the leading hemisphere at the
mid-latitudes. Further, the redundant
Band 6 observations corroborate the cool residual patterns in the
southern, trailing hemisphere and around 
Pwyll crater ($271^\circ$W, $25^\circ$S), which were observed across
observations and at slightly different local Europa time. While the
residual maps from Band 3 appear to correlate somewhat with known
terrain features (Figure \ref{fig:res_map}, panel A), there are few
locations where these residuals are greater than $3\times$ the
background RMS; as such, we note these correlations with caution. Figure
\ref{fig:res_map} (panel C) shows the regions from the averaged Band 6 and 7
map (Figure \ref{fig:res_map}, panel B) where
residual magnitudes are greater than $3\times$ the observation RMS noise
(colored contours) overlaid on a composite image mosaic from the
Voyager 1, 2, and Galileo spacecraft\footnote{https://astrogeology.usgs.gov/search/map/Europa/Voyager-Galileo/Europa$\_$Voyager$\_$GalileoSSI$\_$global$\_$mosaic$\_$500m}. 

 %trailing
Although co-located features exist between the ALMA Band 6 and
7 data both in the best-fit (Figures \ref{fig:res_L}
  and \ref{fig:res_T}) and the global average (Figure
  \ref{fig:res_map}) residuals, these patterns
do not always correlate particularly well with known geographic features or
Europa's albedo distribution. This was previously noted in studies
with the Galileo/PPR \citep{rathbun_10} and ALMA \citep{trumbo_18},
and may result from subsurface properties that vary with the
composition or structure (e.g. crystalline water ice) rather than
macroscale surface terrain. \citet{rathbun_14} found generally higher thermal inertia values on
Europa's trailing hemisphere from Galileo/PPR data, though their
trailing hemisphere values were of lower magnitudes overall than
we find here as inferred through porosity models. A slight difference
was found by \citet{rathbun_14} between chaos and plains regions
across the disk, with the latter requiring slightly lower thermal
inertias, but the correlation with longitude was stronger than with
terrain type; our significantly elevated $\Gamma_{eff}$ values on the
trailing hemisphere corroborate this observation.

Our observations do not show residuals indicative of the
focusing effect of low energy ions and charged
particles on the center of the trailing hemisphere (the colloquially
known `bullseye' pattern seen in models; \citealp{nordheim_22}). In our global average Band 6 and 7 residuals, we find
that Annwn regio (320$^\circ$W, 20$^\circ$N) appears to require lower
thermal inertia (or higher porosity) than the surrounding terrain on
the trailing hemisphere; in contrast, the nearby Dyfed regio
(250$^\circ$W, 10$^\circ$N) shows slightly negative residuals,
requiring higher thermal inertia (lower porosity). The larger area covering these two
regiones has been found to contain signatures of hydrated minerals and 
products of sulfur radiolysis -- whose nature is complicated by the confluence of
endo- and exogenic processes thought to occur at these longitudes --
and dearth of water ice compared to the
leading hemisphere \citep{mcewen_86, carlson_05, grundy_07, brown_13,
  ligier_16, trumbo_20, king_22}. The comparisons of our residual maps
in these longitudes may be somewhat complicated by the
location of the Pwyll crater, which is relatively young and
exhibits extensive rays outward up to $\sim1000$ km
\citep{moore_98, fanale_00, schenk_02, zahnle_03, bierhaus_09}. This
ray system allows the larger extent of Pwyll's ejecta cover an area
equivalent to our
average ALMA beam size ($\sim500$ km at Europa), and is consistently
colder than our Band 6 and 7 models. This may be due its relative
brightness and the ejection of less processed water ice from below,
which has yet to be modified through exogenic processes. Pwyll was evident as a cold residual in the previous
ALMA Band 6 observations \citep{trumbo_17b, trumbo_18}; similarly,
individual large craters and complexes were notably cold in the ALMA
observations of Ganymede and Callisto \citep{de_kleer_21a,
  camarca_23}. The proximity of Pwyll to the nearby regiones thought
to be heavily altered by sulfur radiolysis makes this area
potentially difficult to fit with a single global porosity value,
particularly for moderate ALMA resolution elements compared to the local
features (regiones range from $\sim1500-2500$ km). 

% Leading
We find that the leading hemisphere has larger magnitude residual values, which
similarly indicates a conflict between differing terrain types that
cannot quite be fit by a single, highly porous model. This was found by
\citet{trumbo_18} as well, with the largest range of potential
emissivities and thermal inertia values required to fit longitudes
0--180$^\circ$W. While our most significant positive thermal residuals
are co-located with Tara regio (75$^\circ$W, 10$^\circ$S), the residual
pattern is not confined to it. As the best-fit porosity values for
Europa's leading hemisphere are already elevated compared to the trailing
hemisphere, the large positive residual swath from
$\sim30-90^{\circ}$W potentially represents the highest
porosity (or lowest $\Gamma_{eff}$) or emissivity surfaces we
observe. The lack of significant positive anomalies at
  these locations in the Band 3 data (Figure \ref{fig:res_L}, panel D)
  indicate that these anomalies are not the result of (large)
  endogenic heating, and instead due to compositional or material
  differences that elevate the emissivity or porosity compared to the
  surrounding terrain. Rough or irregular terrain could result in
  elevated temperatures in these regions. Increased salinity (or other non-water
  materials) in the chaos regions could also raise brightness temperatures
  compared to the model through the increase of the complex
  dielectric constant, which in effect would decrease the electrical
  skin depth and reveal more shallow, warmer layers of the regolith. Indeed, recent HST and JWST
observations find NaCl and CO$_2$ to be concentrated in this region
\citep{trumbo_22, trumbo_23, villanueva_23}, thought to be the result
of endogenic emplacement. The
western warm anomalies align somewhat with Powys region, though we do not find a similar cold residual at the location
of the Taliesin crater (138$^\circ$W, 22$^\circ$S) and its surrounding
ejecta blanket as exhibited by Pwyll on the trailing hemisphere. 

We find the coldest ($\Delta T\approx5.5-6$ K) thermal
anomalies at the mid-latitudes of the leading hemisphere. These
locations, while not co-located with known geographic features, align
well with the pure, crystalline water-ice distribution found by
previous studies \citep{hansen_04, brown_13, ligier_16}. Galileo PPR measurements showed warmer nighttime temperatures at
mid-latitudes than the equator on the leading hemisphere, which were
attributed to higher thermal inertia values or endogenic heating \citep{spencer_99,
  rathbun_10}, though \citet{trumbo_18} found a reduced emissivity
(0.66) may be responsible for their cold residual at northern
mid-latitudes. These regions
are impacted by only the highest energy ($\geq1$ MeV) ions and particles from the
Jovian radiation environment \citep{nordheim_22}, and as such have been much less
processed externally than the trailing hemisphere. Thus, it's
possible that the anomalous features we find on the leading hemisphere
are more indicative of the endogenic properties (crystalline water-ice, salts
and carbon-bearing molecules) sourced from Europa's interior
that sculpt its surface. Additionally, as the large thermal
  anomalies exhibited by the Band 6 and 7 data are less significant in the Band
  3 observations, which probe below the thermal skin depth, two
  further possibilities arise: the anomalies present in the Band 6 and 7 data
  are due to thermal inertia variations alone (and thus do not manifest
  at depth), or they are due to thermal
  inertia and emissivity variations that are only substantial down to
  $\sim3$ cm depths. The latter option may occur if the
  variations due to emissivity are not present at the depths probed by
  ALMA Band 3 ($\sim10-20$ cm).

% processes
High energy electrons and their associated bremsstrahlung radiation may still supply the
subsurface with considerable processing down to $\sim10$ cm, while
heavy (S, O)
ions from Io's plasma torus only affect the upper few
millimeters of the surface \citep{paranicas_01, cooper_01, paranicas_02}. While the trailing hemisphere of Europa
receives more total charged particle flux from the Jovian
magnetosphere, the leading hemisphere still receives sufficient dosage
at all but the equatorial latitudes from particles with higher
energies \citep{paranicas_09, nordheim_22};
this, combined with the young relative age of Europa's surface, 
renders the effects of magnetospheric radiation more difficult to
discern on Europa than some of the Saturnian satellites, where the
effects of charged particle bombardment focused on the trailing
hemisphere are more directly evident through thermal emission \citep{howett_14}.
Erosion due to micrometeorite gardening may only affect the regolith
down to 0.5--1 cm \citep{moore_09}, which bounds our Band 7 and 6
measurements (see Appendix \ref{sec:app_b}). While the global average and
best-fit trends between
these frequency bands are fairly consistent, the effects of sputtering
and gardening on regolith grain size and mixing may be important
considerations for
interpreting the best-fit values for our models at different
depths.

  % Caveats/future stuff
Finally, it is worth noting that the average spatial resolution of our ALMA observations is
relatively large compared to various surface features on Europa
(chaos, craters, ringed features; \citealp{doggett_09, leonard_17}), which
warrants future observations at higher angular resolution
(e.g. $\sim100$km or better) to determine
how much the size and shape of the ALMA resolution element affects the morphology of the residual features
we show here. Future observations could also target specific areas at
multiple local Europa times to disentangle the effects of porosity and
emissivity on regional anomalies. Additionally, observations at
additional frequency bands would probe depths above and below those
investigated here, which may allow for better constraints on the
compaction length scale, probe different subsurface processes, and
determine the depth of the anomalous features found in Band 6 and 7. Data
from ALMA bands 8 and 9 (385--500 and 602--720 GHz, respectively) may
be more comparable to Galileo PPR and other IR observations that probe the
shallow subsurface, while polarization measurements may reveal more
about the (sub)surface roughness, scattering, and dielectric properties.

\section{Conclusions} \label{sec:conc}
Through the analysis of multiple ALMA observations of Europa across three
frequency bands -- Band 3 (97.5 GHz; 3.05 mm), Band 6 (233 GHz; 1.25
mm), and Band 7 (343.5 GHz, 0.88 mm), which sound to depths of $\sim0.5$--20 cm -- we are able to infer
the best-fit hemispheric and global average thermophysical properties and brightness
temperatures of the near subsurface at an average resolution of $\sim500$
km. The comparison of the ALMA data to thermal models, incorporating
subsurface emission and
change in surface properties with depth \citep{de_kleer_21a},
allowed for the retrieval of porosity and emissivity values at
discrete depths of $\sim0.5-1$, 1.5--3, and 10--20 cm. Through
these derived properties, we conclude the following:
\begin{itemize}
\item The derived, effective thermal inertia ($\Gamma_{eff}$) values
  of 56--184 J m$^{-2}$ K$^{-1}$ s$^{-1/2}$
  for Europa are within the ranges found previously through studies of
  Voyager and Galileo data at the surface, and comparable to those
  retrieved by \citet{trumbo_18} from separate ALMA Band 6
  observations using different modeling methods.
  
\item Data from ALMA Band 3 revealed colder subsurface temperatures --
  though within the errors of those measured in ALMA Band 6 and 7 --
  that originate from below the thermal skin depth. As a
  result, the thermophysical properties were inferred indirectly, as
  models of subsurface emission for a range of porosity and thermal
  inertia values yielded similar, degenerate fits. The residuals are
  of less statistical significance than those found with the higher frequency ALMA bands.
  
\item Model comparisons with ALMA Band 6 and 7 data show both positive and
negative thermal anomalies
of at least 6$\sigma$, though the total magnitudes are
often \textless5 K. The lowest residuals (\textless1 K) were found for Europa's
trailing antiJovian hemisphere (our image 7T), which is best-fit using
  a global porosity model
(50$\%$) at $\sim1$ cm depths.

\item The derived porosity and brightness temperature values differ
between hemispheres consistently between frequency bands; we find that
Europa's leading hemisphere is generally cooler and more
porous, though large, cold thermal anomalies exist in regions that may
harbor significantly elevated thermal inertia regolith.

\item Despite the differences between leading and trailing
  hemispheres, the best-fit
porosities between bands on the same hemisphere are similar enough
that a compaction length scale cannot be derived. As such, we find no
evidence for large changes in porosity or thermal inertia over the upper
$\sim1-3$ cm. 

\item We find that thermal anomalies only partially align with geographic
features in a consistent way, with larger magnitude positive anomalies co-located with Tara, Powys,
and Annwn regiones, and negative anomalies with Dyfed regio and the vast rays and ejecta of Pwyll crater on the
trailing hemisphere. Negative thermal anomalies on the leading hemisphere -- the
largest we observe of all residual temperatures -- are co-located with
regions of more pure, crystalline water ice. These may be due to elevated
thermal inertia terrain, or a decrease in emissivity that only extends
to \textless10 cm. Positive anomalies
exist in regions with previously observed salt or CO$_2$ features, and to some
extent, chaos regions \citep{leonard_17}.

\item As the depths probed by (sub)millimeter observations are below
  the upper layer of amorphous ice and the regolith
  affected by micrometeorite gardening \citep{hansen_04, moore_09}, our observations
  are more likely sensitive to the distribution of pure, crystalline water
  ice, though warm anomalies may be linked to the mixture of both
  endo- and exogenic processes (e.g. hydrated materials mixed with chaos
  terrain). 

\end{itemize}

ALMA Bands 4 and 5 ($\sim120-230$ GHz) may probe just above the
interface where our Band 3 observations are no longer sensitive to
diurnal variability (just at the thermal skin depth or above), while
ALMA Bands 8 and 9 ($\sim385-720$ GHz) sound the very upper subsurface
($\sim$mm depths), which may provide a means to derive the compaction length
scale of the shallow subsurface and more properly constrain the ALMA
Band 3 measurements. The higher ALMA frequency observations are more
readily comparable to measurements of the surface properties derived
from IR data, while VLA observations at high resolution would probe
\textgreater m depths; data from the \textit{Juno}/MWR will provide
constraints on the thermophysical properties at even greater depths. Tracing thermal emission
from radio to infrared wavelengths will help elucidate the influence of external and internal
processes on Europa's subsurface
properties and structure, and in addition inform our understanding of the surfaces
of other icy satellites.
                
\section{Acknowledgments}
This material is based upon work supported by the
  National Science Foundation under Grant No. 2308280. This research
  was also funded in part by the Heising-Simons Foundation through
  grant $\#$2019-1611. Funding for this paper was provided by the NASA ROSES Solar System
Observations program (through Task Order 80NM0018F0612) for AET, KdK,
AA. Contributions from AA were carried out at the Jet Propulsion
Laboratory, California Institute of Technology, under a contract with
the National Aeronautics and Space Administration (80NM0018D0004). We acknowledge support
from the National Science Foundation Graduate Research Fellowship
under Grant $\#$DGE-1745301 to MC.

We would like to acknowledge
the North American ALMA Science Center staff for their expertise and
help reducing data associated with this project, and in particular to
R. Loomis and L. Barcos-Mu$\tilde{n}$oz for their
assistance during an ALMA face-to-face visit. We would also like to
acknowledge A. Moullet and R. Moreno for their contributions to the original
ALMA proposal on which these observations were based.

This paper makes use of the following ALMA data:
ADS/JAO.ALMA$\#$2016.1.00691.S. ALMA is a partnership of ESO
(representing its member states), NSF (USA) and NINS (Japan), together
with NRC (Canada), MOST and ASIAA (Taiwan), and KASI (Republic of
Korea), in cooperation with the Republic of Chile. The Joint ALMA
Observatory is operated by ESO, AUI/NRAO and NAOJ. The National Radio
Astronomy Observatory is a facility of the National Science Foundation
operated under cooperative agreement by Associated Universities, Inc.

\begin{appendix}
  \section{Removal of Effects of Interloping Moons} \label{sec:app_a}
\begin{figure}
    \centering
    \includegraphics[scale=0.6]{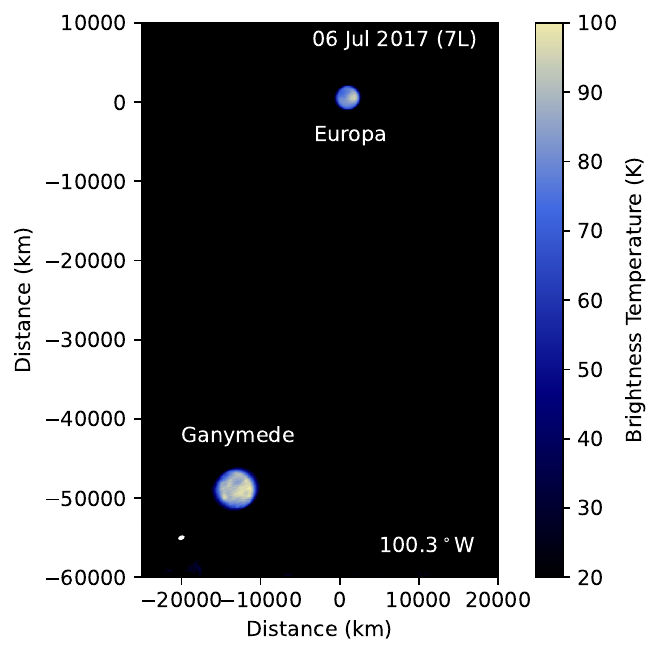}
    \caption{Image targeting Europa's leading hemisphere in Band 7 (top right) with
      Ganymede in the extended ALMA field (lower
      left), approximately $15''$ away. Primary beam correction has
      been applied to normalize the gain across the extended FOV. The
      ALMA beam size appears as the white ellipse in the lower left
      corner.}
    \label{fig:int_im}
  \end{figure}

A noticeable increase in interferometric
artifacts was evident in one observation each in Bands 3 and 7 (3T,
7L; see Table \ref{tab:obs}), resulting in excess background signal
  that was comparable to the thermal modeling residuals. These artifacts were largely removed by accounting for
the presence of Ganymede and Callisto, which were within $\sim40''$ of
Europa during these observations. We achieved this by increasing the
image size from $1000\times1000$ pixels to
\textgreater$4000\times4000$ pixels when creating images. This change
allowed us to include the interloping moon in the image creation and
self-calibration process, significantly reducing the presence of artifacts in the
final image. Figure \ref{fig:int_im}
shows a portion of the larger image created for the leading hemisphere
observation of Europa in Band 7 (7L). The second
satellite, Ganymede, appears in the image on the lower left. The dark region
exhibited on the leading hemisphere of Ganymede was identified as the
Tros impact crater by \citet{de_kleer_21a}, which is similarly cooler
than the disk in their Band 6 images (see their Figure 1). 

\begin{figure}
    \centering
    \includegraphics[scale=0.6]{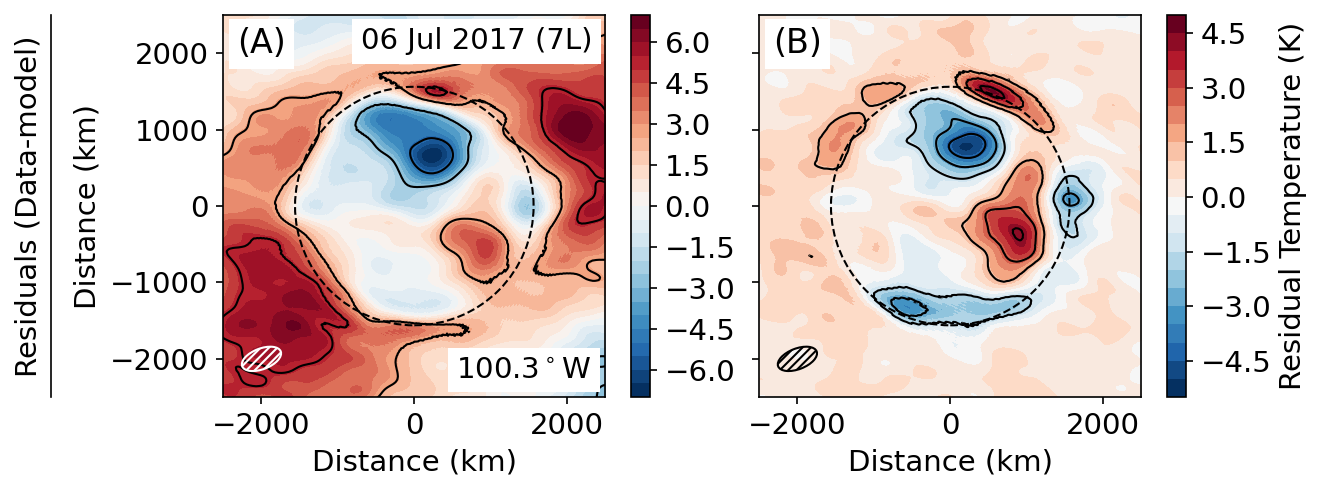}
    \caption{Comparisons of ALMA Band 7 residuals of Europa's leading
      hemisphere: the initial deconvolution, without taking into account
    the interloping satellite (A), and the final deconvolution with
    the inclusion of Ganymede (B); the latter is comparable to Figure \ref{fig:int_im}
    (C). Note the change in colorbar scale between the two
    images. Europa's surface is shown (dashed
    circle), as are separate contours for each image (solid lines): 1$\sigma$
    intervals (A) and 3$\sigma$ intervals (B).}
    \label{fig:b7_res}
  \end{figure}

Figure \ref{fig:b7_res} shows the difference in residual maps
corresponding to imaging performed without the inclusion of Ganymede
(A), and with Ganymede (B), as in Figure \ref{fig:int_im}, using a
nominal global thermal inertia model with $\Gamma=75$ J m$^{-2}$ K$^{-1}$
s$^{-1/2}$. Large, off-disk image artifacts are present in the smaller
image (created without the inclusion of Ganymede; Figure
\ref{fig:b7_res}, A), which are removed when the larger image is
created including Ganymede (Figure \ref{fig:b7_res}, B). Similar artifacts were present in the initial imaging
of the Band 3 trailing hemisphere observation due to the interference
of Callisto. As in Figure \ref{fig:res_L} (C, F), the image in Figure
\ref{fig:b7_res} (B) shows localized thermal anomalies on Europa's
disk only following the inclusion of Ganymede. Following these minor
procedures, a reduction in background signal by factors of $\sim2-4$ were achieved
for these observations -- largely through the decrease in the
background interferometric artifacts. The final image SNR of \textgreater100--200 is much
more comparable to the other observations where interloping satellites
did not affect the data.
  
For future observations of the interior Galilean Satellites -- as well
as those for the Giant Planets -- a careful consideration of the
positions of neighboring satellites should be considered, in addition
to the primary body, when imaging individual satellites.
Other means of removing the effect of nearby planetary bodies in
interferometric observations (e.g. \citealp{de_pater_19a}) can achieve
similar results and may be preferable for different observational
situations. In particular, it is worth noting that the simple approach
employed here is only effective when the
observational duration is short, such that the objects do not move
significantly with respect to one another on the sky. 
  
\section{Derivation of ALMA Band 3 Porosities} \label{sec:app_b}                  
\begin{figure}[h]
  \centering
  \includegraphics[scale=0.7]{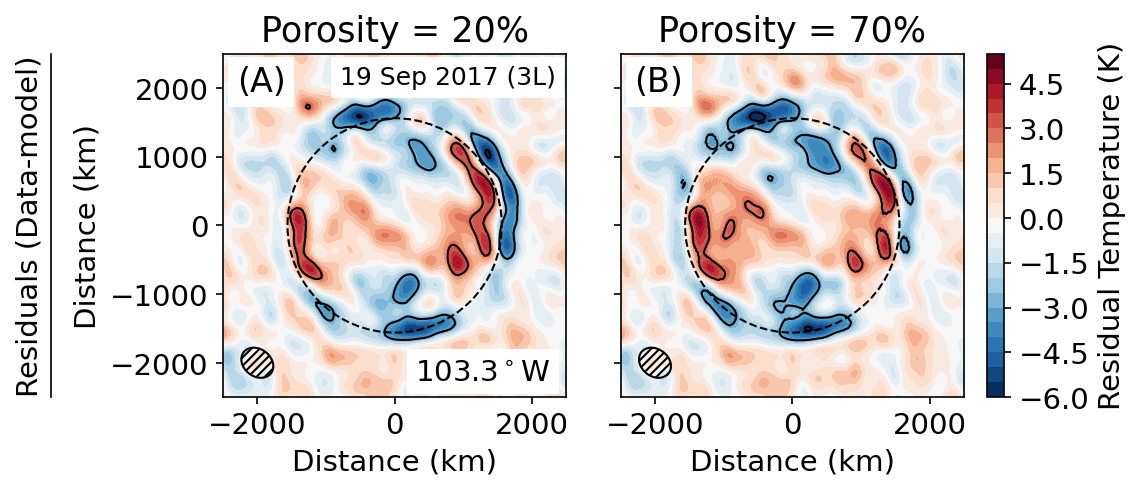}
  \caption{Comparison of residuals for Europa's leading hemisphere in
    ALMA Band 3 using two different global porosity models: 20$\%$
    (A), and 70$\%$ (B). Europa's surface is denoted by a dashed
    circle. Contours, increasing in 3$\sigma$ intervals, are shown
    (solid lines).}
  \label{fig:b3_res}
\end{figure}

Observations 3L and 3T (Figure \ref{fig:res_L} A, D, and Figure
\ref{fig:res_T} A, E, respectively) show low level residuals
when compared to the noise, particularly regions where
thermal anomalies are present at higher frequencies. Further, Figure
\ref{fig:b3_res} shows the comparison of Band 3 residuals following
the subtraction of models with global porosities of 20$\%$ and 80$\%$,
which look remarkably similar. These results are
indicative of thermal emission originating from below the thermal skin
depth ($\delta_T$), where temperature variability due to (sub)surface response to
diurnal fluctuations are no longer substantial. As in
\citet{de_kleer_21a}, this term is parameterized by:

\begin{equation}
  \delta_T = \sqrt{\frac{k_{eff}(p, R, T_{eff})P}{\pi\rho_{eff}(p)c_p(T_{eff})}}
\end{equation}

or alternatively, in terms of the effective thermal inertia,
$\Gamma_{eff}$ (defined in Equation \ref{eq:TI}):

\begin{equation}
 \delta_T = \sqrt{\frac{P}{\pi}}\frac{k_{eff}(p, R, T_{eff})}{\Gamma_{eff}}
 \end{equation}

Here, $P$ is the diurnal period of Europa. For temperatures
relevant to Europa's near surface, $\delta_T$ ranges from
$\sim$5--15 cm depending on porosity or
$\Gamma_{eff}$.

\begin{figure}
  \centering
  \includegraphics[scale=0.5]{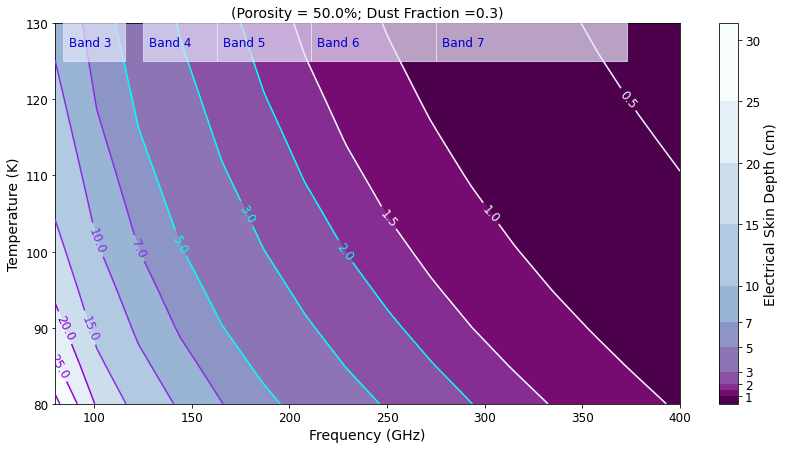}
  \caption{Electrical skin depth as a function of frequency
   in the range of ALMA receiver bands for temperatures relevant to Europa's surface
    and near subsurface for a global 50$\%$ porosity model.}
  \label{fig:d_e}
\end{figure}

This value can be compared to the electrical skin depth, $\delta_E$,
which governs the sensitivity of different wavelengths to thermal
emission vertically throughout the ice crust:
\begin{equation}
  \delta_E = \frac{\lambda}{4\pi\kappa}
\end{equation}
where $\lambda$ is the wavelength, and $\kappa$ is the imaginary
portion of the complex index of refraction, which itself depends on
the ice porosity, dust mass fraction, and temperature (see Section 3
of \citealt{de_kleer_21a}). An example of how $\delta_E$ varies across
notional ALMA frequency bands for a range of temperatures appropriate for
Europa and a surface porosity of 50$\%$ is shown in Figure
\ref{fig:d_e}. These calculations include the multiplicative scale
factor applied to $\kappa$ as discussed in
Section \ref{sec:rad}.
  
\begin{figure}
    \centering
  \includegraphics[scale=0.5]{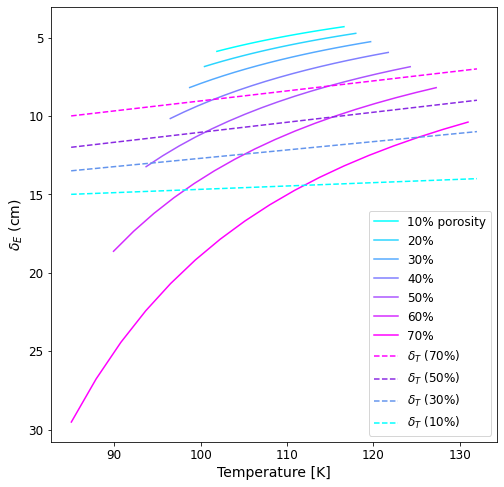}
  \caption{Electrical skin depth ($\delta_E$) curves (solid lines) as
    a function of temperature for a range of global
    porosities at 97.5 GHz
    ($\sim3$ mm), as covered by continuum observations in ALMA Band
    3. Temperature bounds are set by the
    predicted variability Europa's surface experiences
    throughout a nominal diurnal cycle for a given porosity value. The
    dust fraction is set to 0.3.
    Thermal skin depths ($\delta_T$) over the range of temperatures are plotted
    (dashed lines) for 10, 30, 50, and 70$\%$ porosity models, illustrating the
    depths needed for derived $\delta_E$ values to be below
    $\delta_T$, and thus not exhibit temperature anomalies due to
    diurnal variability.}
  \label{fig:de_vs_dt}
\end{figure}

For a range of porosity values and relevant temperatures, the comparison of
both $\delta_T$ and $\delta_E$ (including the applied scale factor, as
in Figure \ref{fig:d_e}) are shown in Figure
\ref{fig:de_vs_dt}. As the porosity of the ice increases, the depth at
which thermal emission may be sensed remotely increases (solid curves
in Figure \ref{fig:de_vs_dt}); conversely, the thermal skin depth
decreases (dashed lines in Figure \ref{fig:de_vs_dt}), and as a result
the diurnal variability influences more shallow layers with higher
porosity. In addition to the aforementioned
parameters, the dust mass fraction alters
the range of depths sounded by radiation -- increasing the dust fraction
decreases $\delta_E$. As such, there exists a parameter space in which
$\delta_E\textgreater\delta_T$, manifesting as residuals with minimal
temperature variability across longitudes and at multiple porosity
values, as we find in ALMA Band 3. Considering permutations of Figure
\ref{fig:d_e} and \ref{fig:de_vs_dt}, we find that a porosity of
$\sim40\%$ marks a physically realistic lower bound for depths down to
$\sim20$ cm, as sounded by ALMA frequencies of $\sim$100 GHz. There does
not exist a combination of parameters for which these data could be sensitive
to emission from the subsurface for porosities lower than $30\%$ while
simultaneously sounding depths below the
thermal skin depth, which would thus manifest more significant thermal
anomalies. Though higher porosity values (e.g. \textgreater$70\%$) allow for sensitivity far
below $\delta_T$, we assume the ice at depth is no more porous than
that of the (near) surface. These physical constraints allow us to
define the bounds for porosities as measured at low frequencies, and thus we infer a porosity of
  $50^{+20\%}_{-10\%}$ or $\Gamma_{eff} = 140^{+43}_{-70}$ for ALMA Band 3, sounding between $\sim8-20$ cm
depending on temperature, porosity, and dust fraction. 

Future observations with ALMA at intermediate frequencies (e.g. ALMA
Band 4 and 5, from $\sim125-211$ GHz) may sound
regions above Band 3 where diurnal temperature variations are still
detectable (Figure \ref{fig:d_e},  allowing for us to further examine the potential porosity
gradient with depth at Europa.

 \end{appendix}

 \pagebreak

\end{document}